\pdfoutput=1
\documentclass[aip,pof,reprint,floatfix,amsmath,amssymb]{revtex4-2}

\usepackage{graphicx}
\usepackage{etoolbox}
\usepackage{float}
\usepackage{amsmath,amssymb,bm}
\usepackage{mathptmx}
\usepackage{booktabs}
\usepackage{xcolor}
\usepackage{hyperref}
\usepackage{url}
\hypersetup{colorlinks=true,linkcolor=blue,citecolor=blue,urlcolor=blue}
\graphicspath{{figures/}}

\newcommand{\Kn}{\mathrm{Kn}}
\newcommand{\Ma}{\mathrm{Ma}}

\newcommand{\relLtwo}{\mathrm{rel}\,L_2}

\newenvironment{widepage}{%
  \clearpage\penalty0\relax
  \global\paperwidth=11in
  \global\paperheight=8.5in
  \global\pdfpagewidth=11in
  \global\pdfpageheight=8.5in
  \begingroup
  \setlength{\textwidth}{10in}%
  \setlength{\textheight}{7in}%
  \setlength{\linewidth}{10in}%
  \setlength{\columnwidth}{10in}%
  \setlength{\hsize}{10in}%
  \setlength{\oddsidemargin}{-0.25in}%
  \setlength{\evensidemargin}{-0.25in}%
  \setlength{\topmargin}{-0.35in}%
}{%
  \clearpage\penalty0\relax
  \endgroup
  \global\paperwidth=8.5in
  \global\paperheight=11in
  \global\pdfpagewidth=8.5in
  \global\pdfpageheight=11in
}

\begin{document}

\title{Bulk-to-Wall Observability in Rarefied Hypersonic Flow: Moment Limits and Incident Half-Range Sufficiency}

\author{Elyas Lekzian}
\affiliation{Department of Aerospace Engineering, Faculty of New Sciences and Technologies, Semnan University, Semnan, Iran}

\author{Ehsan Roohi}
\email[Corresponding author: ]{roohie@umass.edu}
\affiliation{Department of Mechanical and Industrial Engineering, University of Massachusetts Amherst, Amherst, MA 01003, USA}

\date{September 2, 2026}

\begin{abstract}
Wall traction in a rarefied gas is a half-range functional of molecules arriving at and leaving a surface, whereas hybrid solvers, moment methods, and learned wall models often transmit only finite full-range moments. We determine which directional information is lost and what restores the wall functional using direct simulation Monte Carlo (DSMC) of Mach-6 flow over three triangular-protrusion orientations. Output-specialized neural networks interpolate a 57-condition DSMC database. A distinct ExtraTrees regression holds learner capacity, auxiliary features, train/test split, and seeds fixed while comparing the primitive state $S_0$, the momentum-flux-augmented state $S_1$, and the heat-flux-augmented state $S_2$. Adding the momentum-flux tensor $P_{ij}$ reduces aggregate pressure error by 31\% but does not make signed shear transferable. An analytic null-space construction supplies the mechanism: strictly positive distributions can share all full-range moments through degree three while producing different pressure and opposite signed shear because the wall kernel selects the incident half space. A finite-distance off-wall ExtraTrees state $S_{\mathrm{off}}$ shows that partial directionality is useful but representation-dependent. Finally, a parameter-free kinetic reconstruction combines the incident half-range wall-arrival state $S_{\mathrm{HR}}$ with the prescribed diffuse kernel and agrees with both a same-window DSMC tally sharing the incident events and a separate 40,000-step DSMC wall-tally window. The resulting design principle is direct: a rarefied wall closure must preserve the incident normal--tangential correlations required by the target load rather than merely adding higher full-range moments.
\end{abstract}

\maketitle
\onecolumngrid

\section{Introduction}
Hypersonic vehicles rarely present perfectly smooth aerodynamic surfaces. Antennas, sensor windows, wiring ducts, flanges, thermal-protection-system tile offsets, inspection ports, and manufacturing imperfections all behave as localized protrusions that can reorganize the near-wall flow. In continuum hypersonics, isolated roughness elements and protuberances are known to amplify surface heating, alter separation and transition, and produce strong localized pressure loads. Reviews and experiments on hypersonic roughness demonstrate that even geometrically small disturbances can create engineering-scale penalties in heat flux and pressure, especially when they interact with a thin high-speed boundary layer.\cite{Schneider2008,BerryHorvath2008} Step, ramp, pin, and isolated-protuberance studies show that local heating and pressure depend on obstacle geometry, height, Mach number, and upstream boundary-layer state.\cite{Grotowsky2000,Estruch2010,Chang2010,Qamar2012,Hahn2013,KumarReddy2013,KumarReddy2014} A related literature examines roughness-induced instability and transition over plates, cylinders, and capsule-like bodies.\cite{DeTullio2013,Avallone2016,DuanXiao2017,DiGiovanni2018,BlancoCasares2022,Haley2023} Vehicle-scale investigations further demonstrate that pins, distributed protrusions, and launch-vehicle appendages can alter both integrated and asymmetric loads.\cite{Akshay2021,Tsutsui2022,Chandrakumar2023,Leopardi2024}

The rarefied version of this problem is more subtle than a continuum roughness problem. When the mean free path is comparable to the protrusion height, the flow around the obstacle is controlled not only by local gradients but also by molecular flight paths, accommodation at the wall, nonlocal transport, and statistical sampling. Rarefied forward-facing steps, backward-facing steps, gap-step geometries, wall-mounted cubes, and surface-fitted obstacles have therefore been studied with kinetic methods rather than ordinary continuum closure.\cite{PullinHarvey1977,LeiteSantos2015,NabapureMurthy2019,WangFang2020,NabapureMurthy2021,ManelaGibelli2020,Gavasane2018,MahdaviRoohi2022,SabouriLekzian2025} The triangular protrusion database of Sabouri and Lekzian provides a particularly useful benchmark because it isolates three orientations of triangular roughness in a rarefied hypersonic setting.\cite{SabouriLekzian2026} Their study showed that forward-facing protrusions create the strongest in-domain pressure and temperature peaks and the largest surface pressure and heat-transfer coefficients, while backward-facing and isosceles cases generate different vortex structures and surface-load distributions. Those physical trends make the dataset a challenging test of whether a surrogate can learn more than a smooth interpolation of color contours. The orientation- and height-dependent vortex topology was already documented in that direct simulation Monte Carlo (DSMC) study; the present work uses those established flow structures solely to interpret the load footprints and does not repeat a second vortex classification analysis.

The reference solutions considered here are generated by DSMC, a particle method that statistically samples the Boltzmann equation. The method is the standard high-fidelity tool for transition-regime rarefied flows because it does not impose a Navier-Stokes-Fourier closure where the continuum hypothesis is questionable.\cite{Boyd2017,Stefanov2019} The simulations use the SPARTA (Stochastic PArallel Rarefied-gas Time-accurate Analyzer) DSMC solver, which supports locally refined Cartesian and cut-cell meshes for complex boundaries.\cite{Plimpton2019} Accuracy of DSMC data is not automatic: time-step, grid, particle-per-cell, and validation checks are required because the reference field itself contains stochastic scatter. The underlying protrusion simulations used locally refined grids and a fixed-particle-per-cell DSMC strategy, with validation against a hypersonic flat-plate experiment and sensitivity studies for grid, time step, and sampling quality.\cite{Becker1974,Sun2011PPC,Shamseddine2019,Sabouri2024FPPC} The production calculations use monatomic argon with the variable-soft-sphere collision model, fully diffuse isothermal plate and protrusion surfaces, free-stream conditions on the upstream and upper boundaries, and a downstream vacuum outlet. The SPARTA Cartesian cut-cell mesh is locally refined around the protrusion, and the fixed-particle-per-cell implementation maintains approximately 20 simulator particles per active cell; the grid, time-step, and particle-count choices were fixed after the verification studies reported with the source database.\cite{SabouriLekzian2026}

Machine learning can reduce the cost of repeated DSMC queries, but the surrogate must respect the physics and the sampling topology of the data. Physics-informed neural networks and neural operators have made rapid progress in learning parametric maps for fluid and kinetic problems, including rarefied shock structure and shock-aligned surrogate representations.\cite{Raissi2019,Karniadakis2021,Lu2021,Maulik2020,Morimoto2022,Peyvan2026,RoohiMahdavi2026,RoohiShojaAzghadi2026,RoohiMahdaviPoF2026,VinuesaBrunton2022} However, the rarefied protrusion problem differs from smooth canonical operator-learning examples in three ways. First, the mesh is cut-cell and locally refined, so a coordinate-conditioned surrogate is more natural than a uniform-grid convolutional model. Second, the physical outputs have different localization: velocity and temperature are relatively smooth over the gas region, pressure is concentrated in the compression layer, and surface coefficients are one-dimensional wall functions with sharp windward and apex peaks. Third, the database is structured but not fully factorial in all five parameters.

A closely related bulk-to-wall question has a substantial history in continuum wall-bounded turbulence, and the present work should be positioned against it. Wall-modeled large-eddy simulation reconstructs the instantaneous wall stress from resolved outer-layer data through equilibrium or nonequilibrium wall models,\cite{BosePark2018} physics-informed and building-block network wall models learn the same map from data,\cite{Yang2019WM,BaeKoumoutsakos2022,LozanoDuranBae2023} convolutional networks translate in both directions between wall quantities and off-wall velocity fields,\cite{Guastoni2021,Guemes2021} resolvent-based estimators recover space--time flow statistics from limited sensor information,\cite{Towne2020} and data-driven closure modeling in this regime has been reviewed broadly.\cite{Duraisamy2019} All of these efforts operate where a Navier--Stokes--Fourier description holds: the wall stress is then, at least in principle, a functional of the resolved outer state through a thin near-wall layer with exploitable universal structure, and the scientific question is how to model that functional accurately and cheaply. The rarefied transition regime studied here changes the question itself. The wall functional acts on the incident half of the molecular velocity distribution, the near-wall region is a Knudsen layer with no log-law or equilibrium stress balance to lean on, and it is shown constructively below that finite full-range moment states can fail to determine the wall load at all. The continuum literature therefore asks how to model the bulk-to-wall map; this work asks first whether the chosen bulk state makes the wall load observable, and identifies the directional information that must be added when it does not.

Wall traction is a half-space velocity integral of the molecules incident on and emitted from the surface, whereas density, velocity, temperature, stress, and heat flux are full-range moments that mix the two populations. Whether a finite set of such moments determines the wall load is therefore a genuine observability question, and it is not an academic one. Hybrid kinetic--continuum algorithms hand a finite state across a coupling interface and must reconstruct wall fluxes from it; moment methods of the Grad family require wall boundary conditions built from exactly this information;\cite{Grad1949,Struchtrup2005} and learned wall closures and surrogate pipelines are trained to map near-wall fields to loads under the same implicit assumption. In none of these settings is tallying the surface directly an answer, because the kinetic solver whose tallies would be used is precisely what the interface, the closure, or the surrogate is intended to replace at prediction time. The designer must know which class of directional information a reduced near-wall state must retain before the wall functionals can be determined. This issue is examined directly in velocity space with molecular collision statistics and is tested blind across geometry and rarefaction.

The unresolved issue is whether the bulk or off-wall state supplied to a wall model is identifying. This input is defined before the wall-scattering operation; a stress tensor already evaluated on the boundary-constrained wall state would instead report the traction itself. For the predictive bulk-to-wall map, the traction operator acts on the incident half of velocity space, whereas density $\rho$, velocity $\bm u$, temperature $T$, the momentum-flux tensor $P_{ij}$, and the heat-flux vector $q_i$ are full-range integrals that mix the two velocity directions. We therefore test a specific mechanism: finite full-range compression can discard the sign-of-$c_n$ information needed by the wall functional. Completing the second moment should preferentially help normal traction because the pressure kernel has a large quadratic projection, while restoring directional incident information should improve observability when the representation preserves the flux that actually reaches the surface.

We test these predictions with four linked pieces of evidence. First, an equal-capacity blind interpolation test compares the shared parameter--geometry baseline $B_{\mathrm{pg}}$, which contains no local gas-state variables, with the nested states $B_{\mathrm{pg}}\oplus S_0$, $B_{\mathrm{pg}}\oplus S_1$, and $B_{\mathrm{pg}}\oplus S_2$ under the same train/test split. Here $S_0$ is the primitive state, $S_1=S_0\cup\{P_{ij}\}$ adds the complete momentum-flux tensor, and $S_2=S_1\cup\{q_i\}$ further adds heat flux. Second, a constructive velocity-space null redistribution proves that identical retained full-range moments can coexist with different pressure and opposite signed shear, separating information loss from model-capacity failure. Third, a finite-distance incoming state $S_{\mathrm{off}}$ augments $S_2$ with directional descriptors at $0.25$--$2$ local mean free paths and tests whether partial incident information survives away from the wall. Fourth, the wall-arrival incident flux state $S_{\mathrm{HR}}$, together with the prescribed diffuse kernel, provides the upper-information benchmark. Two complementary DSMC comparisons are used for this reconstruction. The same-window, or co-temporal, comparison uses the same sampling interval and the same incident collision events as the reconstruction: the native DSMC tally retains each realized outgoing velocity, whereas the reconstruction replaces that stochastic draw by its conditional mean under the diffuse kernel. It therefore tests estimator implementation and conditional-expectation consistency. The independent-window comparison instead uses a statistically separate 40,000-step DSMC surface tally that shares no collision events with the reconstruction; it tests cross-window repeatability and exposes the reference sampling floor, particularly for weak signed shear. These tests turn the triangular protrusions from a case-specific surrogate exercise into a controlled probe of a transferable physical principle: a wall closure cannot recover a one-sided kinetic functional from a state that has erased the relevant velocity-space directionality. The validated output-specialized surrogates and curated 57-condition database support the controlled comparisons and are documented in Appendix~\ref{app:surrogate}.

\section{Physical problem and data set}
Throughout the kinetic analysis, $\lambda$ denotes the local mean free path and $\delta$ a wall-normal sampling offset.
The problem is a two-dimensional rarefied hypersonic argon flow over a flat plate containing a triangular protrusion. We denote Mach number by $\Ma$ and Knudsen number by $\Kn$. All geometry used by the surrogates and all geometry shown in Fig.~\ref{fig:geometry} are nondimensionalized by the protrusion-base length $h_s$; consequently, the uniform coordinate scaling used in the DSMC input files does not enter the learning variables or the geometric interpretation. The normalized coordinates are
\begin{equation}
 \xi=\frac{x-x_{\rm apex}}{h_s}, \qquad \eta=\frac{y-y_b}{h_s}.
\end{equation}
Here $x_{\rm apex}$ is the streamwise apex coordinate and $y_b$ is the plate/base elevation.
In these apex-centered coordinates, the base endpoints are $(-1.5,0)$ and $(-0.5,0)$ for the backward-facing geometry, $(0.5,0)$ and $(1.5,0)$ for the forward-facing geometry, and $(-0.5,0)$ and $(0.5,0)$ for the isosceles geometry; the apex is $(0,h_p/h_s)$ in every case, where $h_p$ is the protrusion height. Hence the normalized base length is unity, while orientation is represented solely by the apex position relative to the base. This definition is used consistently in solid masks, distance features, field coordinates, and wall-profile coordinates and prevents the bounding-box extrema from being misidentified as the physical base. Throughout the paper the three orientations are abbreviated ISO for the isosceles protrusion, BWD for the backward-facing ramp whose long shallow face is windward, and FWD for the forward-facing ramp whose steep face is windward.

The physical regime targeted by the data is worth stating at the outset. All conditions place the protrusion inside a cold hypersonic stream of monatomic argon over an isothermal flat plate, and the sampled Knudsen numbers span the transition regime between near-continuum and strongly rarefied behavior. In this window the compression layer over the windward face is several local mean free paths thick, velocity slip and temperature jump at the surface are order-one fractions of their near-wall scales, and the leeward face can fall partly into the ballistic shadow of the apex, so no single continuum correlation covers the sampled space. Triangular protrusions were chosen as the geometric family because one parameterization spans the three canonical responses of a surface irregularity, a symmetric obstruction together with its forward- and backward-leaning limits, while remaining simple enough that every wall element has an unambiguous face assignment. The apex-centered normalization serves the same purpose from the data side: it places the three orientations and the height sweep on one geometric footing, so fields and wall profiles from different cases can be compared bin by bin without interpolation onto a common physical mesh. These regime features are not incidental background. They are what make the wall-load question kinetic rather than hydrodynamic, and they set the scales, mean free path, slip length, and shadow extent, against which every later comparison in the paper is judged.

\clearpage
\begin{widepage}
\begin{figure}[H]
\centering
\includegraphics[width=0.98\linewidth]{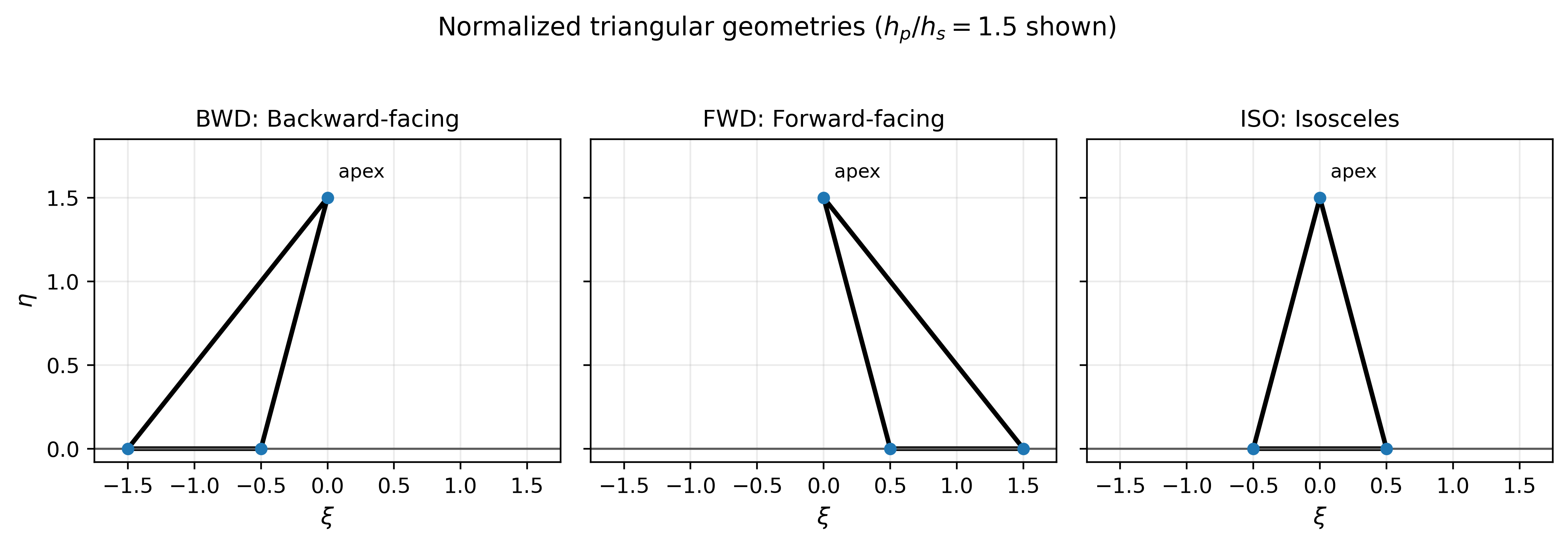}
\caption{Normalized triangular geometries used in this work; $h_p/h_s=1.5$ is shown for illustration. The axes are $\xi=(x-x_{\rm apex})/h_s$ and $\eta=(y-y_b)/h_s$, so the numerical coordinate scaling used in the DSMC input files does not affect the geometry supplied to the surrogates. The base length is unity in these coordinates.}
\label{fig:geometry}
\end{figure}
\end{widepage}

The data of our simulations contains 69 raw DSMC files: 27 files in the complete Mach--rarefaction--orientation block, 24 files in the height sweep, and 18 files in the wall-temperature sweep. Twelve files repeat baseline physical conditions already present in the complete block: six $h_p/h_s=1.5$, $T_w/T_\infty=1$ cases in the height sweep and six $T_w/T_\infty=1$, $h_p/h_s=1.5$ cases in the wall-temperature sweep. Removing these exact duplicate physical keys leaves 57 unique conditions, equivalently $27+18$ additional-height conditions at $h_p/h_s=1.0,1.25,$ and $2.0$, plus 12 additional wall-temperature conditions at $T_w/T_\infty=2$ and $4$. The cases span Mach numbers $4$, $6$, and $8$; Knudsen numbers $0.1$, $0.33$, and $0.8$ in the Mach-rarefaction-orientation block; all three protrusion orientations; height ratios $h_p/h_s$ from $1.0$ to $2.0$ at Mach number $6$; and wall-temperature ratios $T_w/T_\infty$ from $1$ to $4$ at Mach number $6$. Table~\ref{tab:topology} states this topology explicitly. The first block is a full factorial sampling of Mach number, Knudsen number, and orientation at fixed height ratio and wall temperature. The height-sweep block tests geometric sensitivity at Mach number $6$ and wall-temperature ratio $1$. The wall-temperature block tests thermal sensitivity at Mach number $6$ and height ratio $1.5$. This design supports interpolation tests along the sampled axes; simultaneous nonbaseline height--temperature, height--Mach, and wall-temperature--Mach interactions lie outside the sampled topology. A provenance package accompanies the manuscript and lists all 69 raw keys, the 57 deduplicated physical keys, and duplicate mappings. A compact summary of the available run-level particle counts, processor allocations, reported runtimes, and normalized time steps is provided in Appendix~\ref{app:dsmc_execution}; the complete record is supplied as a comma-separated-values (CSV) file. Execution-accounting metadata are available for 51 of the 57 unique conditions; the six omitted records are the $h_p/h_s=2.0$ height-sweep cases, whose field and surface data remain part of the 57-condition database.

Three design choices behind this data deserve a brief physical justification, because they control what every later test can and cannot claim. The one-factor sweeps around a shared baseline were chosen so that each interpolation direction isolates a single mechanism: the Mach sweep changes the strength of the compression layer at fixed collisionality, the Knudsen sweep changes collisionality at fixed forcing, the height sweep changes the obstruction geometry, and the wall-temperature sweep changes only the emitted population through the kernel temperature. Deduplication matters for the same reason. Twelve raw files repeat physical conditions that already exist under other names, and if one physical state appeared on both sides of a learning split, every blind error in this paper would be optimistically biased; collapsing the 69 raw keys to 57 unique physical conditions before any split is what makes the word blind meaningful. The execution-accounting record, finally, ties each condition to its particle count, processor allocation, and time step, so that the statistical weight of every data entry can be audited rather than assumed.

\begin{table}[t]
\centering
\caption{Supported sampling topology of the DSMC database. The validation and interpretation in this paper are restricted to these sampled directions.}
\label{tab:topology}
\small
\begin{ruledtabular}
\begin{tabular}{llll}
Sampling block & Varied quantities & Fixed quantities & Supported sensitivity \\
Mach--rarefaction--orientation & \parbox[t]{0.25\textwidth}{Mach number $\{4,6,8\}$; Knudsen number $\{0.1,0.33,0.8\}$; ISO/FWD/BWD orientation} & \parbox[t]{0.18\textwidth}{$h_p/h_s=1.5$; $T_w/T_\infty=1$} & \parbox[t]{0.25\textwidth}{Full factorial Mach number $\times$ Knudsen number $\times$ orientation.} \\
Height sweep & \parbox[t]{0.25\textwidth}{$h_p/h_s\in\{1.0,1.25,1.5,2.0\}$; Knudsen number $\{0.1,0.33\}$; orientation} & \parbox[t]{0.18\textwidth}{Mach number $6$; $T_w/T_\infty=1$} & \parbox[t]{0.25\textwidth}{Height sensitivity and height-ratio $\times$ Knudsen-number variation at Mach number $6$.} \\
Wall-temperature sweep & \parbox[t]{0.25\textwidth}{$T_w/T_\infty\in\{1,2,4\}$; Knudsen number $\{0.1,0.33\}$; orientation} & \parbox[t]{0.18\textwidth}{Mach number $6$; $h_p/h_s=1.5$} & \parbox[t]{0.25\textwidth}{Wall-temperature sensitivity and wall-temperature $\times$ Knudsen-number variation at Mach number $6$.} \\
\end{tabular}
\end{ruledtabular}
\end{table}

\subsection{Case and evidence manifest}
Table~\ref{tab:evidence_manifest} identifies the computational class of every contribution. Neural networks emulate the 57-condition database; ExtraTrees regression provides the controlled tests of $S_0$--$S_2$ and $S_{\mathrm{off}}$; the remaining components are the analytic null-space construction and the parameter-free kinetic reconstruction from $S_{\mathrm{HR}}$. The complete case mapping is supplied in \texttt{data/manifest\_raw.csv}, \texttt{data/manifest\_unique\_physical\_conditions.csv}, and the SPARTA pilot manifests. Exact duplicate physical keys are used only once in learning and validation splits. Collision archives are reserved for the kinetic reconstruction, while every empirical prediction is scored against direct DSMC surface impulses.

\begin{table}[h]
\centering
\caption{One-glance map of the DSMC evidence, neural-network emulation, ExtraTrees observability tests, analytic null-space construction, and kinetic event reconstruction. Here $B_{\mathrm{pg}}$ denotes the shared parameter--geometry baseline, which contains no local gas-state variables.}
\label{tab:evidence_manifest}
\footnotesize
\setlength{\tabcolsep}{3pt}
\begin{tabular}{p{0.18\textwidth}p{0.12\textwidth}p{0.24\textwidth}p{0.18\textwidth}p{0.19\textwidth}}
\toprule
Evidence block & Cases/input & Operation & Computational class & Reference/target \\
\midrule
DSMC database & 69 files / 57 unique & Field and wall emulation & Neural network & DSMC fields and tallies \\
Moment endpoints & 6 & Train $B_{\mathrm{pg}}$, $B_{\mathrm{pg}}\oplus S_0$, $B_{\mathrm{pg}}\oplus S_1$, $B_{\mathrm{pg}}\oplus S_2$ & ExtraTrees regression & DSMC wall impulses \\
Moment interpolation & 6 & Blind test of the same four inputs & ExtraTrees regression & DSMC wall impulses \\
Finite-distance pilot & 4 & Blind $S_{\mathrm{off}}$ test & ExtraTrees regression & DSMC wall impulses \\
Null-space construction & Positive velocity pairs & Moment nonuniqueness & Analytic construction & Exact moment map \\
Half-range collision pilot & 6 & $S_{\mathrm{HR}}$ reconstruction & Kinetic event sum & DSMC wall tally \\
Independent wall reference & Same 6 & Cross-window repeatability & DSMC sampling & Four time blocks \\
\bottomrule
\end{tabular}
\end{table}

The held-out cases test interpolation in physically meaningful directions. The first validation group removes the Mach number $6$, Knudsen number $0.33$ slice across all orientations from the full Mach-rarefaction-orientation block. The second removes the intermediate height ratio $h_p/h_s=1.25$ in the height-sweep block. The third removes the intermediate wall-temperature ratio $T_w/T_\infty=2$ in the wall-temperature block. Together these tests evaluate supported sensitivities: the coupled effect of Mach number, rarefaction, and orientation; height dependence at Mach number $6$; and wall-temperature dependence at Mach number $6$.

The kinetic-sufficiency study uses a controlled SPARTA pilot at Mach number 6. Production endpoints use $\Kn=0.1$ and 0.8 for all three orientations; blind interpolation cases use $\Kn=0.2$ and 0.4. Each case contains four statistically separated field and wall blocks. The field data of every pilot case stores the primitive fields together with the complete momentum-flux tensor components $P_{xx}$, $P_{yy}$, $P_{zz}$, and $P_{xy}$, the translational heat-flux vector, and block-resolved wall impulses. Surface targets are accumulated from molecular momentum exchange, not inferred from the nearby cell pressure. Separate continuation runs record pre-collision velocities on the protrusion faces in five 1000-step blocks. The pressure and signed-shear comparisons use 60 arclength bins and Student-$t$ 95\% confidence intervals. The same-window (co-temporal) DSMC tally is accumulated over the collision-recording interval and therefore shares the realized incident events with the half-range reconstruction. An independent reference is formed from four separate 10,000-step surface-tally blocks (40,000 steps total) and shares no collision events with the reconstruction. This design provides two distinct checks: same-window conditional-expectation consistency and independent-window repeatability under DSMC noise. A separate off-wall extraction provides incoming molecular descriptors at normalized wall-normal offsets $\delta/\lambda=0.25$, $0.5$, $1$, and $2$ for the BWD and FWD endpoint-trained/intermediate-$\Kn$ cases. These data are used only in the four-case $S_{\mathrm{off}}$ pilot. Five model seeds are retained, and the direct DSMC wall impulses remain the targets; no wall-arrival collision event is supplied to the learned finite-distance state.

\section{Output-specialized surrogate and kinetic-information framework}
\begin{figure}[H]
\centering
\includegraphics[width=\textwidth,height=0.44\textheight,keepaspectratio]{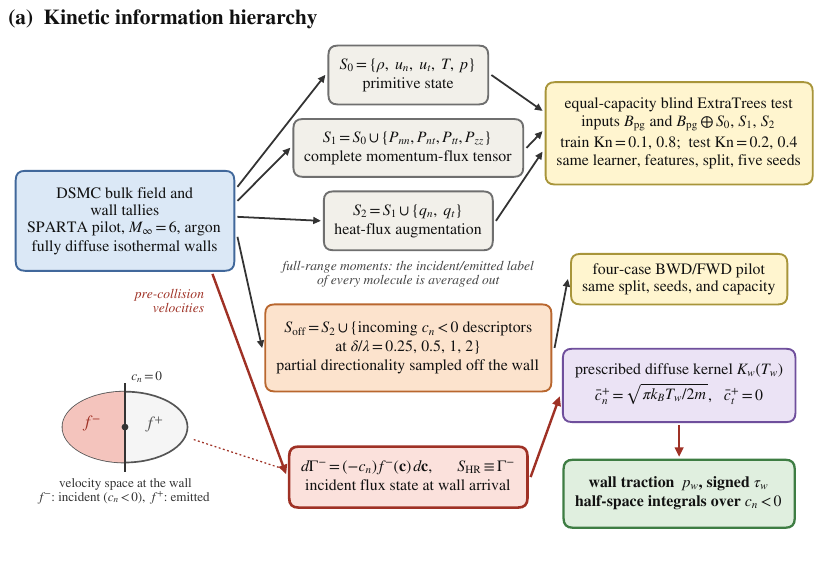}\\[3mm]
\includegraphics[width=\textwidth,height=0.44\textheight,keepaspectratio]{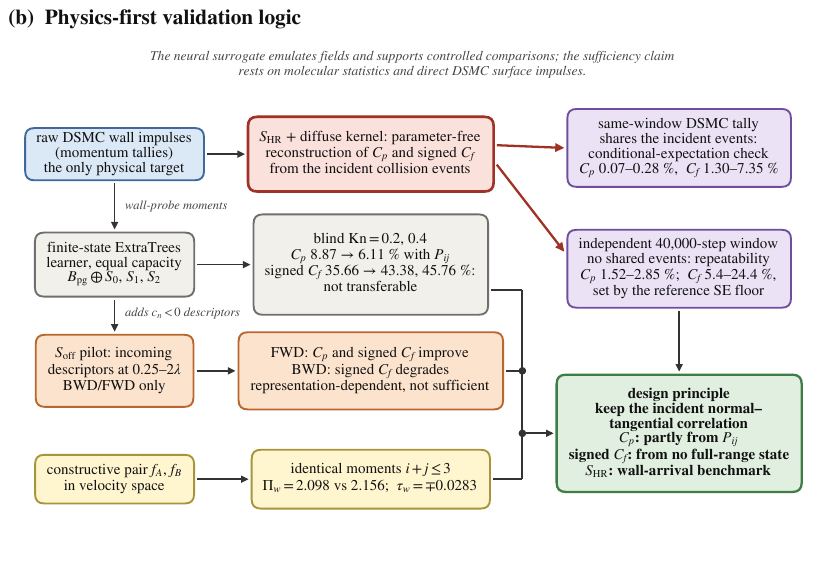}
\caption{Kinetic information hierarchy and validation logic. (a) $S_0$, $S_1$, and $S_2$ are nested finite full-range moment states; $S_{\mathrm{off}}$ adds partial incident directionality, and $S_{\mathrm{HR}}$ is the wall-arrival incident flux state. (b) Direct DSMC surface impulses define every physical target. ExtraTrees regression evaluates $S_0$--$S_2$ and $S_{\mathrm{off}}$; the velocity-space pair supplies the analytic null-space test; and the parameter-free $S_{\mathrm{HR}}$ event sum is checked against a same-window DSMC tally sharing the incident events and a statistically separate 40,000-step DSMC tally.}
\label{fig:framework}
\end{figure}

The framework is organized by information content rather than by network architecture. Figure~\ref{fig:framework}(a) separates three finite full-range states from the incident half-range state. In local wall-normal and wall-tangent coordinates,
\begin{align}
S_0 &= \{\rho,u_n,u_t,T,p\},\\
S_1 &= S_0\cup\{P_{nn},P_{nt},P_{tt},P_{zz}\},\\
S_2 &= S_1\cup\{q_n,q_t\},
\end{align}
where $P_{ij}=\int m c_i c_j f\,d\bm{c}$ and $q_i=\int \tfrac12m c^2c_i f\,d\bm{c}$ are full-range moments, $m$ is molecular mass, $\bm c$ is molecular velocity, and $f$ is the velocity distribution. The finite-distance directional state augments $S_2$ with incoming-only descriptors at four normal offsets; the subscript ``off'' denotes off-wall sampling,
\begin{equation}
S_{\mathrm{off}}=S_2\cup\{\text{incoming descriptors at }\delta/\lambda=0.25,0.5,1,2\}.
\label{eq:soffstate}
\end{equation}
At each offset particles are projected into the local wall-normal/tangent frame and only $c_n<0$ particles enter the directional descriptors. The compact state includes normalized incoming number content and flux-weighted combinations of normal, tangential, and energy moments. It therefore restores some incident/emitted separation without using the wall-arrival collision events themselves. The BWD/FWD pilot is trained at $\Kn=0.1$ and $0.8$ and tested blindly at $\Kn=0.2$ and $0.4$, with five seeds.

For the Table~\ref{tab:momentgate} ExtraTrees comparison, $S_0$, $S_1$, and $S_2$ are evaluated at wall-normal probes $\delta/\lambda=0.5$, $1$, and $2$. The common parameter--geometry baseline $B_{\mathrm{pg}}$ contains the wall position and height-normalized coordinate, the local tangent and gas-facing normal, and $\log_{10}\Kn$, but no local gas-state variables. Primitive variables are normalized by their free-stream values, $P_{ij}$ by $p_\infty$, and $q_i$ by $p_\infty U_\infty$; nearest-cell miss distance normalized by $\lambda$ is retained at every probe. The four tested inputs are $B_{\mathrm{pg}}$ alone, $B_{\mathrm{pg}}\oplus S_0$, $B_{\mathrm{pg}}\oplus S_1$, and $B_{\mathrm{pg}}\oplus S_2$. Unused state columns are filled with their training-set means so all four regressors have identical input dimension and capacity.

By contrast, let $\bm n$ point from the wall into the gas and let $\bm t$ define the positive local tangent. With $c_n=\bm c\cdot\bm n<0$ for an incident molecule, the collision records sample the wall-arrival measure
\begin{equation}
d\Gamma^{-}(\bm c)=(-c_n)f^{-}(\bm c)\,d\bm c,\qquad S_{\mathrm{HR}}\equiv\Gamma^{-}.
\label{eq:shrstate}
\end{equation}
Thus $S_{\mathrm{HR}}$ is the incident flux state seen by the wall, with HR denoting half range, not another finite full-range moment set. For a specified gas--surface scattering kernel $K_w$, let $\langle\bm c^{+}\rangle_{K_w|\bm c}$ denote the mean velocity emitted after an incident molecule with velocity $\bm c$. The force per unit area exerted by the gas on the wall is
\begin{equation}
\bm{t}_w=\int_{c_n<0}m\left[\bm c-\langle\bm c^{+}\rangle_{K_w|\bm c}\right]d\Gamma^{-}(\bm c),
\qquad p_w=-\bm t_w\cdot\bm n,\quad \tau_w=\bm t_w\cdot\bm t.
\label{eq:walltraction}
\end{equation}
Equation~(\ref{eq:walltraction}) is the continuous mapping used throughout the half-range part of the study. It maps the incident flux state of Eq.~(\ref{eq:shrstate}) and the known wall kernel to the two wall targets. The normalized coefficients are $C_p=(p_w-p_\infty)/(\tfrac12\rho_\infty U_\infty^2)$ and $C_f=\tau_w/(\tfrac12\rho_\infty U_\infty^2)$.

Figure~\ref{fig:framework}(b) shows the validation logic. Raw DSMC surface impulses are the target. Full-range states are compared with identical model capacity and blind intermediate-Knudsen-number holdouts. $S_{\mathrm{off}}$ uses a separate four-case BWD/FWD pilot to ask whether partial incident directionality sampled away from the surface is already useful. The wall-arrival $S_{\mathrm{HR}}$ result is computed directly from collision statistics and the known diffuse-wall kernel, then compared first with the same-window (co-temporal) DSMC tally and subsequently with the statistically separate 40,000-step DSMC window. The neural surrogate remains valuable for rapid field reconstruction and as a controlled comparison tool, but it does not define the kinetic sufficiency claim. The three comparisons belong to different inference classes: $S_0$--$S_2$ and $S_{\mathrm{off}}$ are learned finite states, whereas $S_{\mathrm{HR}}$ is a parameter-free wall-arrival oracle evaluated through the prescribed kernel.

The discrete estimator clarifies the distinction between the incident-state reconstruction and the native DSMC wall tally. Let a wall element of area $A$ collect, during a collision block of duration $\tau$, pre-collision velocities $\bm c_k$ from the SPARTA collision-event tally. Each record is an impact event and therefore already samples the arrival measure $d\Gamma^{-}$. If $m$ is molecular mass and $w$ is the simulator-particle weight, the incident-only Monte Carlo estimator is
\begin{equation}
p_w=\frac{w}{A\tau}\sum_k m\left(\overline c_n^{+}-c_{n,k}\right),
\qquad
\tau_w=\frac{w}{A\tau}\sum_k m\left(c_{t,k}-\overline c_t^{+}\right).
\label{eq:reconinc}
\end{equation}
Equation~(\ref{eq:reconinc}) defines the incident-state estimator. Its shear term uses each realized incoming tangential velocity $c_{t,k}^{-}$ and the analytic kernel mean $\overline c_t^{+}$ for the emitted state. The native DSMC tally in Eq.~(\ref{eq:dsmcrealized}) uses the realized pair $c_{t,k}^{-}-c_{t,k}^{+}$ for every collision. Thus the two estimators share incoming events and differ exactly in their treatment of the stochastic outgoing draw.
For a stationary fully diffuse wall, the flux-averaged emitted velocity is fixed analytically by the wall kernel,
\begin{equation}
\overline c_n^{+}=\sqrt{\frac{\pi k_B T_w}{2m}},\qquad \overline c_t^{+}=0.
\label{eq:reconemit}
\end{equation}
Here $k_B$ is the Boltzmann constant and $T_w$ is the wall temperature; the molecular mass $m$ is the same as in Eq.~(\ref{eq:reconinc}).
The relation to the realized DSMC wall tally is made explicit because the two quantities are deliberately related estimators, not independent physical observables. For the same incoming events, let $\bm c_k^{-}$ and $\bm c_k^{+}$ denote the realized pre- and post-collision velocities. Then
\begin{equation}
\bm t_{\mathrm{DSMC}}=\frac{w}{A\tau}\sum_k m\left(\bm c_k^{-}-\bm c_k^{+}\right),
\label{eq:dsmcrealized}
\end{equation}
whereas the incident-state reconstruction replaces the stochastic outgoing draw by its conditional expectation under the prescribed wall kernel,
\begin{equation}
\bm t_{\mathrm{HR}}=\frac{w}{A\tau}\sum_k m\left[\bm c_k^{-}-\mathbb{E}\!\left(\bm c^{+}\mid \bm c_k^{-},K_w\right)\right].
\label{eq:hrconditional}
\end{equation}
Consequently,
\begin{equation}
\bm t_{\mathrm{HR}}=\mathbb{E}\!\left[\bm t_{\mathrm{DSMC}}\mid\{\bm c_k^{-}\},K_w\right]
\label{eq:rao-blackwell}
\end{equation}
for the prescribed scattering process. Thus $\bm t_{\mathrm{HR}}$ is the conditional-expectation, or Rao--Blackwellized, counterpart of the stochastic wall tally and both estimators have the same expectation. The terms ``same-window (co-temporal)'' and ``independent-window'' refer to the statistical relationship between the reconstruction and the DSMC reference. In the same-window comparison, both estimators use the same incoming collision events and sampling interval; they differ only because the native DSMC tally retains each stochastic outgoing velocity while the half-range reconstruction substitutes its analytic conditional mean. The resulting error is therefore an implementation and finite-sample consistency check, not an independent-data test. In the independent-window comparison, the reference wall tally is accumulated over four separate 10,000-step blocks, 40,000 steps in total, and shares no impact events with the reconstruction. Differences in this comparison contain sampling variability from both windows. For signed shear, the reference uncertainty reported in Table~\ref{tab:halfrange} is the standard error of the independent DSMC reference profile divided by the RMS magnitude of that profile and expressed as a percentage; weak-shear cases can consequently have large normalized uncertainty even when their absolute fluctuations are small.

Equations~(\ref{eq:reconinc}) and (\ref{eq:reconemit}) are the event-sum implementation of Eq.~(\ref{eq:walltraction}) and the formulas used in the supplied analysis code. Surface identifier and pre-collision velocity are read for every impact. The velocity is projected onto the gas-facing normal and positive tangent, the impulse is summed element by element and block by block, and the result is divided by $A\tau$. The element values are normalized with the same free-stream dynamic pressure used for the DSMC surface tally. Figure~\ref{fig:wallvalidation} shows representative 60-bin profiles, and Table~\ref{tab:halfrange} reports all six cases against same-window and independent-window DSMC references. The reconstruction inputs are the incident collision records and prescribed wall kernel; DSMC wall tallies provide the validation references.

As the top rung of the information ladder, $S_{\mathrm{HR}}$ establishes, on the same geometry, wall bins, and DSMC noise, the precision attainable when the incident arrival state is retained. The combined observability evidence consists of the blind finite-state ExtraTrees test, the analytic full-range-moment null-space construction, the finite-distance $S_{\mathrm{off}}$ ExtraTrees pilot, and the wall-arrival kinetic benchmark. Together they separate inference from finite macroscopic states, fundamental directional information loss, transport of a compact directional representation, and the wall-arrival limit under the known kernel.

The ordering of the finite states is deliberate and forms an information ladder rather than an arbitrary feature list. $S_0$ carries what any hydrodynamic description transports. $S_1$ completes the entire second velocity moment, so that everything a Navier--Stokes--Fourier or regularized-moment closure could build from stress information is available without modeling error. $S_2$ adds the third-degree flux content that enters Grad-type hierarchies through the heat flux. Each rung therefore corresponds to a definite physical statement, that the wall load is a function of transported quantities up to that tensorial order, and the blind test is designed so that every rung is falsifiable on the same data with the same learner capacity, the same features, the same splits, and the same seeds. Two further choices protect the comparison. The parameter-and-geometry baseline, which sees only case labels and wall coordinates and no field information at all, sets the floor any moment state must beat before a claim of useful bulk content can be made. The finite-distance $S_{\mathrm{off}}$ rung asks whether adding directional content before wall arrival is enough, while the half-range reconstruction sits at the top as a zero-parameter conditional-expectation benchmark rather than a learned map. The gap among these rungs therefore separates representation loss from optimization failure and from the distinct problem of transporting an off-wall directional state to the surface.

For the finite-state test, every model is a two-output ExtraTrees regressor for $(C_p,C_f)$ with 800 trees, minimum leaf size 2, and 0.85 feature fraction.\cite{Geurts2006} The five seeds are 11, 29, 47, 71, and 101. Training uses the near-field wall samples of the six endpoint cases, all three orientations at $\Kn=0.1$ and 0.8; testing uses the 60 protrusion-wall bins of each of six blind cases at $\Kn=0.2$ and 0.4. For each seed, the normalized root-mean-square error (NRMSE) is computed after concatenating the six complete test profiles, and Table~\ref{tab:momentgate} reports the mean over seeds. The 95\% interval for the $S_1$-over-$S_0$ pressure gain is a case bootstrap applied to the seed-averaged predictions.\cite{Efron1979} This finite-state split is distinct from the six-case collision pilot described in Table~\ref{tab:evidence_manifest}. The executable feature extraction and training scripts, case-by-seed scores, and aggregate record are supplied with the reproducibility package.

For the half-range reconstruction, every recorded incoming simulator particle contributes its normal and tangent momentum flux to its wall element and time block. The emitted contribution is the analytic expectation of the fully diffuse reflection kernel at the prescribed wall temperature. Counts are accumulated before arclength binning so that empty or weakly sampled elements cannot be silently interpolated. Agreement is quantified by the normalized root-mean-square error (NRMSE), the root-mean-square (RMS) of the binwise difference between reconstruction and reference divided by the RMS of the reference over the same bins and reported in percent, together with the correlation coefficient of the binned profiles. Lines in the profile figures use a seven-point, second-order Savitzky--Golay filter separately on the two faces solely for display; raw binned markers, confidence intervals, NRMSE values, and correlations are always computed from the unsmoothed arrays, and no smoothing crosses the apex.

The output-specialized surrogates that establish database-level field and wall interpolation, together with their architecture, error metrics, full validation, region-conditioned errors, engineering quantities of interest, and the thermodynamic-consistency diagnostic, are documented in Appendix~\ref{app:surrogate}. The main text quotes only their headline accuracies and proceeds directly to the kinetic question.

\clearpage
\clearpage

\section{Kinetic information required for wall traction}
\subsection{Blind test of finite full-range moments}
The equal-capacity intermediate-$\Kn$ test provides the first direct answer. Table~\ref{tab:momentgate} reports pooled-profile NRMSE over six blind geometry--rarefaction cases, averaged over five seeds. Adding $P_{ij}$ to the primitive state reduces the aggregate $C_p$ error from 8.87\% to 6.11\%, a relative gain of 31.15\% with a case-bootstrap interval of 3.83--43.68\%. The gain is physically consistent with pressure being a normal momentum flux, but it is not uniform across the test set. The change is negligible for BWD at $\Kn=0.2$, adverse for BWD at $\Kn=0.4$, and favorable for the FWD and ISO cases. The parameter--geometry baseline $B_{\mathrm{pg}}$, which contains no local flow-state variables, remains better at 3.20\%, while adding $q_i$ produces only a small further aggregate change. Signed shear behaves more decisively. $S_1$ and $S_2$ increase the error from 35.66\% for $S_0$ to 43.38\% and 45.76\%, respectively. Thus, for these blind intermediate states, the complete full-range stress and heat-flux moments are neither sufficient nor reliably beneficial for wall shear.

\begin{table}[t]
\centering
\caption{Equal-capacity ExtraTrees test of the incremental wall-load information supplied by finite full-range moment states under blind interpolation from $\Kn=0.1,0.8$ to $\Kn=0.2,0.4$. Values are pooled six-case wall-profile NRMSE (\%), averaged over five seeds. The parameter--geometry baseline $B_{\mathrm{pg}}$ contains the wall location and height-normalized coordinate, local tangent and gas-facing normal, $\log_{10}\Kn$, and probe miss-distance information, but no gas-state variables. The column $B_{\mathrm{pg}}\oplus S_0$ adds the primitive state $S_0=\{\rho,u_n,u_t,T,p\}$ at $\delta/\lambda=0.5,1,2$; $B_{\mathrm{pg}}\oplus S_1$ additionally supplies the complete momentum-flux tensor $P_{ij}$; and $B_{\mathrm{pg}}\oplus S_2$ further adds the heat-flux vector $q_i$. Model capacity, auxiliary features, training/testing cases, and random seeds are identical in all columns.}
\label{tab:momentgate}
\begin{tabular}{lcccc}
\toprule
Target & $B_{\mathrm{pg}}$ only
& $B_{\mathrm{pg}}\oplus S_0$
& $B_{\mathrm{pg}}\oplus S_1$
& $B_{\mathrm{pg}}\oplus S_2$ \\
\midrule
$C_p$ & 3.20 & 8.87 & 6.11 & 5.84 \\
Signed $C_f$ & 26.37 & 35.66 & 43.38 & 45.76 \\
\bottomrule
\end{tabular}
\end{table}

The result cannot be explained away solely as insufficient regressor capacity; the deeper reason is kinematic and lives in velocity space.

\subsection{Why finite full-range moments can lose the wall information}
In the local frame associated with a target wall element, the bulk/interface distribution separates into two velocity directions. The incident half-range $f^{-}$ contains molecules with $\bm{c}\cdot\bm{n}<0$ traveling toward the surface; the complementary half has $\bm{c}\cdot\bm{n}>0$. Once the gas--surface interaction is prescribed, the emitted wall population is generated from $f^{-}$ by the scattering kernel, $f^{+}=K_w[f^{-}]$. For the fully diffuse isothermal wall, the emitted state is a wall-temperature Maxwellian whose amplitude is fixed by the incident mass flux, so the wall traction of Eq.~(\ref{eq:walltraction}) is a functional of $f^{-}$. Written with the indicator $\Theta(-c_n)$ and $\overline c_n^+=\sqrt{\pi k_BT_w/(2m)}$, the diffuse-wall pressure and shear functionals read
\begin{equation}
p_w=\int \Theta(-c_n)\,m(-c_n)(\overline c_n^+-c_n)f\,d\bm{c},\qquad
\tau_w=\int \Theta(-c_n)\,m(-c_n)c_t f\,d\bm{c},
\label{eq:halfrangefunctionals}
\end{equation}
where the reflected normal contribution is contained in $\overline c_n^+$ and the reflected tangential mean is zero. Full-range moments, by contrast, are integrals of smooth polynomials over the whole velocity space. No finite polynomial basis can represent these one-sided weights, so for any finite full-range hierarchy there exist non-negative redistributions of $f$ that are invisible to every retained moment yet transfer flux-weighted probability across the plane $c_n=0$. This is the classical reason why half-range or two-stream representations were introduced for wall-bounded kinetic problems,\cite{GrossZiering1958,Cercignani1988} and why moment closures in the Grad family require separately constructed half-flux wall boundary conditions instead of an evaluation of the bulk closure at the surface.\cite{Grad1949,Struchtrup2005} Near equilibrium the difficulty is hidden: a near-Maxwellian distribution is essentially fixed by its low-order moments, and the half-range integrals collapse to ordinary functions of $\rho$, $\bm{u}$, $T$, and the transport fluxes. In a Knudsen layer beneath a Mach 6 stream nothing of the sort holds. The local state is bimodal, combining a cold, fast, incident beam with a hot, slow, diffusely re-emitted population, and low-order full-range moments average over exactly the directional distinction that the wall functionals in Eq.~(\ref{eq:halfrangefunctionals}) need.

A constructive discrete-velocity example, Fig.~\ref{fig:nonunique}, makes the statement quantitative, and the object at its center deserves a careful definition because the whole argument rests on it. In plain terms, a null redistribution is a way of rearranging molecules in velocity space that no retained moment can detect. Formally, collect the ten full-range moments of total degree at most three into one linear moment map, $\mathcal{M}[f]=\{\langle v_n^{i}v_t^{j},f\rangle:\,i+j\le3\}$, where the bracket denotes the sum over the discrete velocity grid; a perturbation $\delta(v_n,v_t)$ is null when it lies in the kernel of this map, $\mathcal{M}[\delta]=0$. The physical reading of the ten conditions is direct. Such a rearrangement carries no mass, no momentum in either component, and no energy, since $v_n^{2}$ and $v_t^{2}$ both belong to the retained set; it alters no entry of the pressure tensor and no third-degree flux moment, including both components of the heat flux. Every collisional invariant of the gas, and every quantity that $S_0$, $S_1$, and $S_2$ report, is left exactly where it was.

The construction can be read as an inverse-problem statement. The retained moments are a finite list of weighted averages of the velocity distribution. Knowing those averages constrains the distribution but does not identify it: many different arrangements of probability among velocity cells give the same list. A null redistribution is the difference between two such arrangements. Starting from a positive base distribution $f_0$, we choose a nonzero $\delta$ with $\mathcal{M}[\delta]=0$ and scale it so that
\begin{equation}
f_A=f_0-\delta>0,\qquad f_B=f_0+\delta>0.
\label{eq:nullpair}
\end{equation}
Linearity then gives $\mathcal{M}[f_A]=\mathcal{M}[f_B]$ exactly. A model receiving only these moments must assign the same input state to $f_A$ and $f_B$, irrespective of its architecture or training quality.

The wall distinguishes the pair because it does not average both velocity directions with the same weight. It samples only $v_n<0$ and weights each incident molecule by its arrival rate and momentum transfer. Therefore a redistribution may remove probability from one incident-velocity band, place it in another band, and compensate on the complementary half so that every full-range polynomial average cancels. The cancellation is exact for $S_0$--$S_2$, yet it is absent from the half-range wall integral. This is the operational meaning of nonuniqueness here: identical retained inputs correspond to different admissible wall loads.

Figure~\ref{fig:nonunique} displays the construction in the $(v_n,v_t)$ plane for visual clarity. Its three-velocity extension is obtained by multiplying both members by the same normalized, even distribution $g(v_z)$, so that $f_{A,B}^{(3V)}=f_{A,B}(v_n,v_t)g(v_z)$. Every three-dimensional monomial through total degree three then factors into an identical $(v_n,v_t)$ moment and a common $v_z$ moment. Consequently $P_{zz}$ and all components of the three-dimensional heat flux remain identical as well, while the pressure and tangential-traction functionals retain the same difference shown in the figure.

The amount of hidden freedom is easy to count. On the $41\times41$ grid the moment map imposes ten linear conditions on $1681$ unknowns, so its kernel has dimension $1671$. Retaining the full hierarchy through degree three therefore pins down less than one percent of the directions available to the discrete state. The only question of substance is whether any of the remaining directions couples to the wall.

Symmetry answers that question exactly, and this is the precise content of the figure. Split each wall weight of Eq.~(\ref{eq:halfrangefunctionals}) into a polynomial half and a discontinuous remainder. Writing $a=\overline c_n^+$ in thermal units gives
\begin{equation}
\begin{aligned}
\Theta(-v_n)(v_n^2-a v_n)&=\tfrac12(v_n^2-a v_n)-\tfrac12\mathrm{sign}(v_n)(v_n^2-a v_n),\\
\Theta(-v_n)(-v_nv_t)&=-\tfrac12v_nv_t+\tfrac12\mathrm{sign}(v_n)v_nv_t.
\end{aligned}
\label{eq:weightsplit}
\end{equation}
A null $\delta$ cannot see the polynomial halves, by definition, so its entire effect on the wall is carried by the discontinuous remainders,
$\Pi_w[\delta]=-\tfrac{1}{2}\langle \mathrm{sign}(v_n)(v_n^2-a v_n),\delta\rangle$ and
$\tau_w[\delta]=\tfrac{1}{2}\langle \mathrm{sign}(v_n)v_nv_t,\delta\rangle$.
Reading the two overlaps gives simple selection rules. A null redistribution shifts the incident pressure exactly when it contains an odd-in-$v_n$ component with structure finer than cubic; it shifts the incident shear exactly when it contains a component even in $v_n$ and odd in $v_t$; every other direction in the kernel is wall-silent.

The same decomposition explains why pressure benefits before signed shear in the blind DSMC hierarchy. The pressure weight contains a low-order polynomial projection, $\tfrac12(v_n^2-a v_n)$, that is represented directly once the second-moment state is complete; the sign-weighted remainder is not. Signed shear is more demanding. Although $P_{nt}$ contains the polynomial term $-\tfrac12 v_nv_t$, the wall also requires $\tfrac12\operatorname{sign}(v_n)v_nv_t$, an incident directional correlation that full-range $P_{nt}$ does not determine. A redistribution can therefore compensate between the incident and emitted halves, leave the complete full-range cross moment unchanged, and still reverse the wall-shear sign. The $C_p$/$C_f$ contrast is consequently a statement about how strongly each wall kernel projects onto the retained moment space, not a generic claim that one regression target is simply easier to learn.

The two pairs in the figure are built directly from these rules. A smooth oscillatory generator with the required symmetry, $\sin(2.4\,v_n)$ for the pressure pair and $\cos(2.2\,v_n)\tanh(1.1\,v_t)$ for the shear pair, is orthogonalized against all polynomials of total degree three or less in the $f_0$-weighted inner product and then multiplied by the bimodal base state $f_0$ itself. The resulting $\delta$ is smooth, decays wherever $f_0$ decays, and never exceeds forty-five percent of $f_0$ pointwise, so both pair members $f_{A}=f_0-\delta$ and $f_{B}=f_0+\delta$ stay strictly positive everywhere; a final least-squares projection onto the discrete kernel removes the floating-point residue of the weighted construction. Within each pair, all ten full-range moments then agree to within $1.4\times10^{-16}$, which is machine precision for the computation.

The diffuse-wall pressure functionals of the first pair are nevertheless $\Pi_w[f_A]=2.098$ and $\Pi_w[f_B]=2.156$ in thermal units, a difference of 2.7\%, and the second pair carries wall-shear functionals of $\mp0.0283$, equal in magnitude and opposite in sign. The third column of the figure exposes the mechanism. For the pressure pair, the null redistribution moves probability between bands of normal velocity on the incident side while compensating elsewhere so that every full-range moment is untouched. For the shear pair, the redistribution has a lobed structure, odd in the tangential velocity, that reverses the correlation between normal and tangential motion only for molecules flying toward the wall. The fourth column integrates the corresponding flux densities and shows how one full-range moment state supports two different one-sided integrals. Consequently, even exact knowledge of this finite full-range hierarchy does not determine the wall functional, and no regressor consuming only $S_0$--$S_2$ can be guaranteed to recover signed shear because the sign is not uniquely determined by its inputs. The example establishes pointwise nonuniqueness; the blind DSMC test above demonstrates its practical relevance in the sampled flow family.

It is instructive to place the construction next to the classical closure viewpoint. A Grad-type ansatz represents the distribution as a Maxwellian multiplied by a polynomial in the retained moments, and within that ansatz the moment-to-distribution map is invertible by construction, so wall fluxes are determined; this is exactly why such closures succeed near equilibrium.\cite{Grad1949,Struchtrup2005} The counterexample lives in the complement of the ansatz. The null redistribution is, by definition, orthogonal to every retained polynomial direction, so it parameterizes precisely the degrees of freedom the ansatz discards, and the blind test shows that at Mach 6 in the transition regime those discarded directions carry wall-relevant content. The construction adds that they can carry it with either sign.

\clearpage
\begin{widepage}
\begin{figure}[H]
\centering
\includegraphics[width=0.96\linewidth]{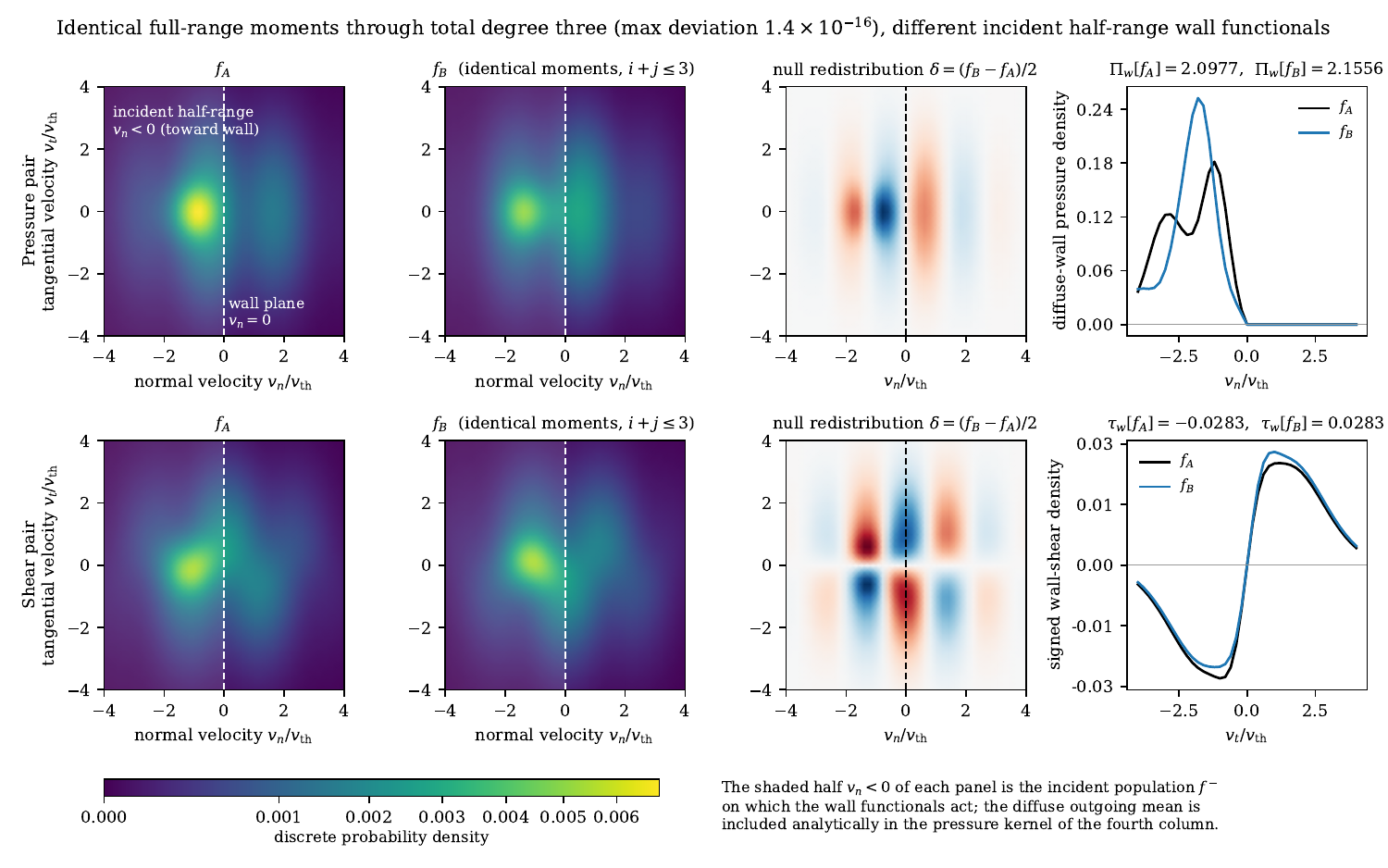}
\caption{Analytic constructive nonuniqueness of finite full-range moments in velocity space. Each row pairs two strictly positive discrete distributions $f_A$ and $f_B$ (first two columns) that share every full-range monomial moment $\langle v_n^{i}v_t^{j}\rangle$ with $i+j\le3$ to within $1.4\times10^{-16}$. The dashed line is the incident/emitted divider $v_n=0$; the shaded half $v_n<0$ is the incident population on which the traction functionals act. The third column shows the null redistribution $\delta=(f_B-f_A)/2$. The fourth column uses the diffuse-wall pressure and signed-shear kernels from Eq.~(\ref{eq:halfrangefunctionals}). Their one-sided integrals give $\Pi_w=2.098$ versus $2.156$ for the pressure pair and $\tau_w=\mp0.0283$ for the shear pair.}
\label{fig:nonunique}
\end{figure}
\end{widepage}

\subsection{Finite-distance incoming information: partial recovery is not sufficiency}
The constructive example identifies the missing variable as velocity-space directionality, but the wall-arrival state is intentionally information rich. A practical intermediate question is whether some of that directionality is already useful before the molecules reach the surface. Table~\ref{tab:offwallpilot} compares $S_0$, $S_1$, $S_2$, and $S_{\mathrm{off}}$ on the four blind BWD/FWD intermediate-$\Kn$ cases for which off-wall incoming samples are available. The same endpoint training split, intermediate-$\Kn$ test split, and five-seed protocol are retained.

The result is deliberately not a monotone ``more information is always better'' story. For FWD, $S_{\mathrm{off}}$ reduces $C_p$ NRMSE from 6.83\% to 6.06\% at $\Kn=0.2$ and from 6.06\% to 4.91\% at $\Kn=0.4$ relative to $S_2$. Signed-shear error also falls from 59.49\% to 49.15\% and from 40.07\% to 37.39\%, respectively. In BWD, however, the pressure change is small while signed-shear error rises from 40.54\% to 42.11\% at $\Kn=0.2$ and from 35.26\% to 42.61\% at $\Kn=0.4$. Finite-distance directional sampling is therefore informative but not universally sufficient.

\begin{table}[H]
\centering
\caption{ExtraTrees regression finite-distance observability test for the BWD and FWD cases. Values are profile NRMSE (\%). All models are trained only at $\Kn=0.1$ and $0.8$ and tested at $\Kn=0.2$ and $0.4$. $S_{\mathrm{off}}$ augments $S_2$ with incoming descriptors at $0.25$, $0.5$, $1$, and $2$ local mean free paths; learner capacity, split, and seeds remain unchanged.}
\label{tab:offwallpilot}
\begin{tabular}{llrrrr}
\toprule
Geometry & Target & $S_0$ & $S_1$ & $S_2$ & $S_{\mathrm{off}}$ \\
\midrule
BWD, $\Kn=0.2$ & $C_p$ & 8.39 & 6.08 & 7.56 & 7.43 \\
 & signed $C_f$ & 28.57 & 36.07 & 40.54 & 42.11 \\
BWD, $\Kn=0.4$ & $C_p$ & 8.50 & 10.82 & 9.00 & 8.46 \\
 & signed $C_f$ & 25.23 & 27.10 & 35.26 & 42.61 \\
FWD, $\Kn=0.2$ & $C_p$ & 8.56 & 11.29 & 6.83 & 6.06 \\
 & signed $C_f$ & 73.81 & 62.92 & 59.49 & 49.15 \\
FWD, $\Kn=0.4$ & $C_p$ & 7.99 & 9.72 & 6.06 & 4.91 \\
 & signed $C_f$ & 60.33 & 47.36 & 40.07 & 37.39 \\
\bottomrule
\end{tabular}
\end{table}

The opposite BWD/FWD response should not be promoted to a universal orientation law. Local geometry changes how an off-wall sample is connected to the wall-arrival population. For an incident direction $\widehat{\bm c}$ sampled a normal distance $\delta$ from a face, the collisionless path required to reach that face scales as
\begin{equation}
\ell_{\rm path}/\lambda\sim \frac{\delta/\lambda}{|\widehat{\bm c}\cdot\bm n|}.
\label{eq:pathscale}
\end{equation}
Near-normal incidence therefore preserves a more direct geometric connection between an offset sample and the arrival state than oblique incidence, while collisions and lateral displacement can decorrelate a compressed off-wall descriptor from the eventual wall flux. The present four-case pilot does not fit Eq.~(\ref{eq:pathscale}) as a scaling law. Its significance is more limited and more useful: partial incident directionality can survive away from the wall, but the representation and its transport must preserve the particular arrival correlation required by the target functional. The BWD shear degradation is retained because it falsifies the simpler claim that any directional descriptor sampled within a few mean free paths is equivalent to the wall-arrival state.

\subsection{Incident half-range sufficiency and independent-window repeatability}
The wall-arrival test covers six geometry--rarefaction cases: ISO at $\Kn=0.1$ and $0.8$, and BWD and FWD at $\Kn=0.2$ and $0.4$. For every case, Eqs.~(\ref{eq:reconinc}) and (\ref{eq:reconemit}) act on incident collision events and the prescribed diffuse kernel. Table~\ref{tab:halfrange} reports the complete six-case comparison against both the co-temporal wall tally and the independently sampled 40,000-step wall tally.

Co-temporal pressure NRMSE is $0.07$--$0.28\%$, and signed-shear NRMSE is $1.30$--$7.35\%$. The independent-window pressure comparison remains at $1.52$--$2.85\%$. Independent signed-shear differences follow the reference signal-to-noise ratio: BWD carries the strongest tangential load and remains at $5.52$--$6.77\%$, while the relative values increase for the weak-shear ISO and FWD cases in proportion to their independently measured reference uncertainty.

\begin{table}[t]
\centering
\caption{Parameter-free incident half-range reconstruction compared with same-window and statistically independent DSMC wall tallies. ``Same window (co-temporal)'' denotes a comparison in which the reconstruction and native DSMC tally share the realized incident collision events; the reconstruction replaces only the stochastic outgoing velocity by its conditional mean under the prescribed diffuse-reflection kernel. ``Independent window'' denotes comparison with a separate 40,000-step DSMC surface tally sharing no collision events with the reconstruction. Entries are wall-profile NRMSE (\%). The final column reports the standard error (SE) of the independent signed-$C_f$ reference profile normalized by the RMS magnitude of that reference profile (\%); large values therefore identify weak-shear, noise-sensitive references.}
\label{tab:halfrange}
\footnotesize
\begin{tabular}{lrrrrr}
\toprule
Case & \shortstack{$C_p$ NRMSE\\same window} & \shortstack{$C_p$ NRMSE\\independent window} & \shortstack{Signed $C_f$ NRMSE\\same window} & \shortstack{Signed $C_f$ NRMSE\\independent window} & \shortstack{Independent signed-$C_f$\\reference SE/RMS} \\
\midrule
ISO, $\Kn=0.1$ & 0.28 & 2.85 & 7.35 & 17.13 & 18.76 \\
ISO, $\Kn=0.8$ & 0.16 & 1.96 & 2.08 & 5.42 & 6.28 \\
BWD, $\Kn=0.2$ & 0.25 & 2.41 & 1.95 & 6.77 & 6.23 \\
BWD, $\Kn=0.4$ & 0.15 & 1.70 & 1.30 & 5.52 & 4.24 \\
FWD, $\Kn=0.2$ & 0.08 & 1.56 & 3.07 & 24.41 & 20.70 \\
FWD, $\Kn=0.4$ & 0.07 & 1.52 & 1.47 & 11.82 & 13.93 \\
\bottomrule
\end{tabular}
\end{table}

Figure~\ref{fig:wallvalidation} presents two complementary cases: BWD at $\Kn=0.2$ for strong, well-resolved shear and FWD at $\Kn=0.4$ for weak shear under near-normal incidence. Pressure and signed shear are shown together, with the kinetic reconstruction, same-window DSMC tally, and independent-window DSMC tally superposed. The table retains the complete six-case evidence, and the elementwise arrays for every case remain in the reproducibility package.

\clearpage
\begin{widepage}
\begin{figure}[H]
\centering
\includegraphics[width=0.98\linewidth]{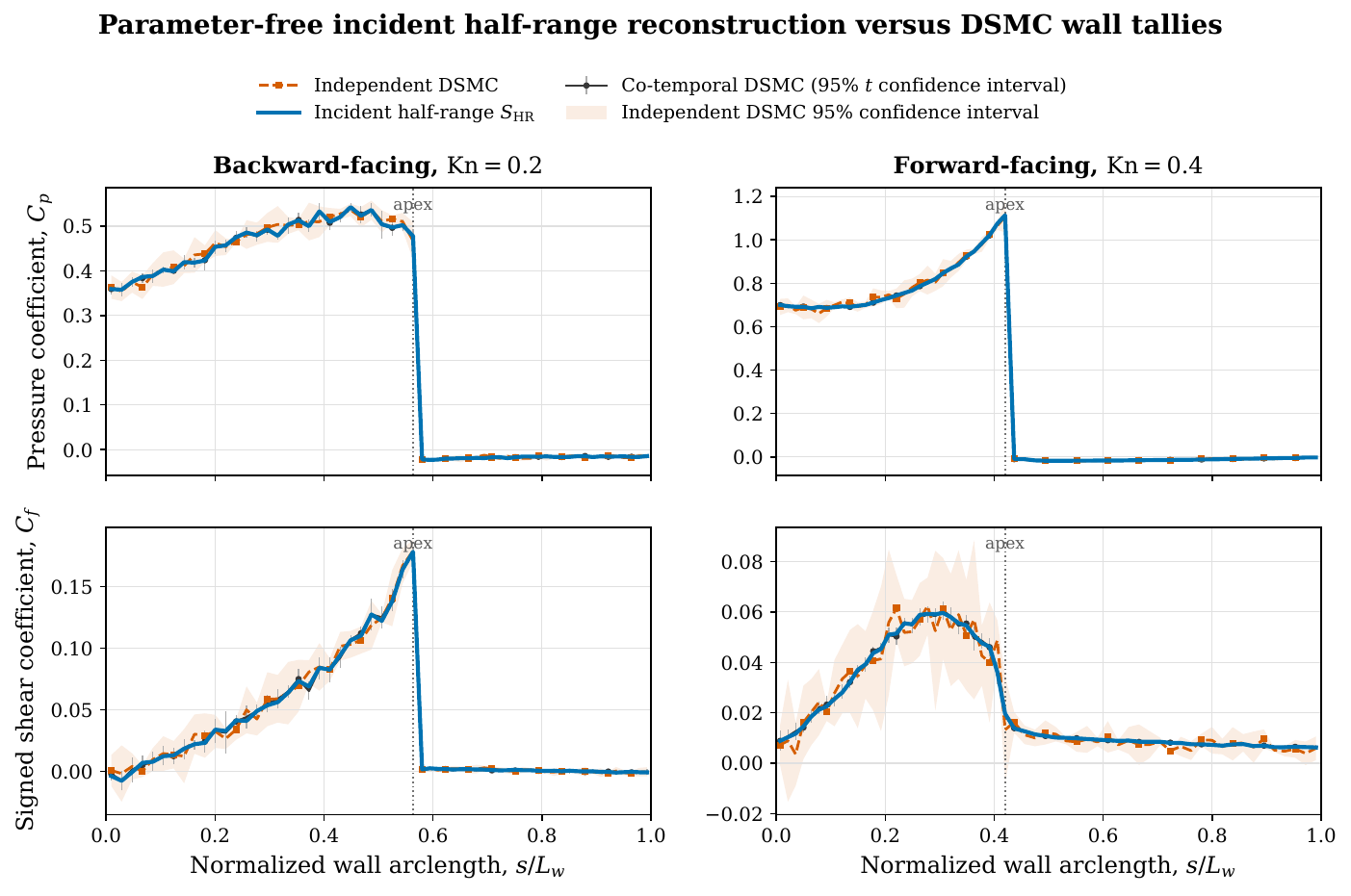}
\caption{Representative wall profiles for a strong-shear BWD case and a weak-shear FWD case. The $S_{\mathrm{HR}}$ curve is the parameter-free kinetic reconstruction from pre-collision velocities and the analytic diffuse-wall mean. Co-temporal DSMC points use realized incoming--outgoing impulses; their vertical bars are two-sided 95\% Student-$t$ intervals. Dashed curves and shaded bands are the independent 40,000-step DSMC reference and its two-sided 95\% interval. The dotted line marks the apex, and Table~\ref{tab:halfrange} reports all six cases.}
\label{fig:wallvalidation}
\end{figure}
\end{widepage}

\section{Physical interpretation and implications}
The central physical result is not a hierarchy of machine-learning scores. It is a decomposition of the wall-load problem into information that a finite full-range state can and cannot carry. Pressure and shear kernels act only on molecules approaching the wall. Full-range moments remove the incident/emitted label, and a finite polynomial hierarchy can therefore preserve mass, momentum, energy, stress, and heat flux while changing a one-sided wall functional. The constructive null-space pair proves that possibility without reference to a regressor. The blind DSMC test shows that the possibility matters in the sampled Mach-6 Knudsen layers. The $S_{\mathrm{off}}$ experiment then shows that restoring directionality away from the surface is beneficial only when the chosen representation preserves the particular incident correlations needed by the load. Finally, $S_{\mathrm{HR}}$ quantifies the accuracy attainable when no directional compression is imposed at wall arrival.

This interpretation explains the quantity dependence. Normal traction overlaps strongly with second-order velocity content, so $P_{ij}$ can supply useful pressure information even when it does not uniquely identify the half-space integral. Signed shear is more fragile because it is an incident correlation between normal arrival and tangential momentum. The full-range cross moment $P_{nt}$ permits compensation between incoming and outgoing populations and therefore does not protect the sign of the wall shear. The relevant control is consequently the projection of the wall kernel onto the retained state together with the unresolved sign-weighted half-space remainder. This is a kinetic statement that survives a change of learner architecture.

Geometry enters by rotating the incident velocity distribution relative to the local wall frame. On the steep FWD windward face, impacts are closer to normal: pressure is large while signed shear is a smaller residual. On the shallow BWD windward face, a larger tangential component survives to impact and shear becomes stronger. These different projections are consistent with the different utility of $S_{\mathrm{off}}$, but the present pilot does not establish a universal FWD--BWD law or a universal scalar ``information length.'' It shows instead that the useful directional representation depends on both the wall functional and how the incident distribution is projected and transported by the local geometry.

The transferable design principle is therefore explicit. A kinetic--continuum interface, moment wall boundary condition, or learned wall model should be designed around the incoming half-space functionals it must reproduce, not around the availability of convenient full-range fields. Candidate reduced states include half-range quadratures, directional moments, or low-dimensional encodings constrained to preserve incident mass, normal momentum, tangential momentum, and energy flux. Their quality can now be tested against two separate benchmarks: $S_{\mathrm{off}}$ asks whether a deployable finite-distance representation transports enough information to the wall, while $S_{\mathrm{HR}}$ sets the wall-arrival limit under a specified scattering kernel.

The consistency of the database-level validation trends with the known protrusion physics, and the ablation record behind the surrogate design choices, are collected in Appendix~\ref{app:surrogate}.

\section{Scope of validity}
The strongest conclusion is an information statement for two-dimensional monatomic argon with a known fully diffuse isothermal wall kernel. The constructive example establishes nonuniqueness for full-range moments through degree three at a point and illustrates the broader mismatch between finite polynomial moment spaces and the discontinuous half-space selector. The blind DSMC hierarchy demonstrates practical non-identification within the sampled protrusion family; it does not prove that every spatially resolved finite-moment field is non-identifying in every kinetic flow.

$S_{\mathrm{off}}$ is an intermediate pilot, not a universal closure. It is presently available only for BWD and FWD endpoint-trained/intermediate-$\Kn$ tests, and its compact incoming descriptors at four distances are one representation among many. The improvement in FWD and the degradation of BWD shear therefore establish representation dependence, not a universal orientation law or an optimum distance. A stronger mechanistic test would condition the off-wall representation on molecular flight-to-wall histories or expand the directional basis until the BWD shear deficit converges.

$S_{\mathrm{HR}}$ is sampled at the wall and uses the same diffuse kernel as the DSMC calculation. Its co-temporal agreement is a conditional-expectation positive control; it is not an independent wall model. The independent 40,000-step reference breaks shared-sampling correlation and resolves pressure robustly, while the weakest signed-shear cases remain limited by the reference noise. The completed 50,000-step FWD collision extension reduces co-temporal shear NRMSE to 3.07\% at $\Kn=0.2$ and 1.47\% at $\Kn=0.4$, but it cannot reduce noise in the earlier independent wall reference. Extending the conclusion to Maxwell or Cercignani--Lampis scattering requires a kernel sweep, and extending it to internal molecular modes, chemistry, three-dimensional roughness, or different Mach numbers requires new kinetic data.

The sampled height and wall-temperature sweeps remain restricted to Mach number 6 and do not include simultaneous nonbaseline combinations. The entropy quantity remains a local-equilibrium compatibility diagnostic for the database-level analysis.

\section{Conclusions}
This study asks when a near-wall rarefied state contains enough information to determine wall traction. The answer is controlled by velocity-space directionality rather than by surrogate capacity alone.

First, the blind full-range hierarchy shows a quantity-dependent limit. Adding the complete momentum-flux tensor to primitive variables reduces aggregate $C_p$ error by 31\%, consistent with the quadratic projection of the pressure kernel, but neither stress nor heat flux makes signed shear transferable. Second, the constructive velocity-space pair identifies the cause: a redistribution can be invisible to every retained full-range moment through degree three while changing diffuse-wall pressure and reversing incident shear. The missing information is the sign-weighted half-space remainder of the wall functional.

Third, the finite-distance $S_{\mathrm{off}}$ experiment tests that mechanism away from the wall. Incoming descriptors at $0.25$--$2\lambda$ improve both pressure and signed shear for the FWD cases, including a reduction of $C_p$ NRMSE to 4.91\% at $\Kn=0.4$, but they do not improve BWD shear. Partial directionality is therefore useful but not generically sufficient; the representation must preserve the particular arrival correlations required by the target functional. Fourth, $S_{\mathrm{HR}}$ combined with the known diffuse kernel gives the upper benchmark. Its co-temporal agreement with the DSMC tally is explicitly a conditional-expectation positive control, while the independent window establishes repeatability at the DSMC noise level. The completed FWD collision continuation confirms co-temporal signed-shear NRMSE of 3.07\% at $\Kn=0.2$ and 1.47\% at $\Kn=0.4$.

The transferable physical principle is that a rarefied wall model cannot be judged only by how accurately it reconstructs nearby macroscopic fields. Its interface state must preserve the incident directional fluxes to which the gas--surface kernel responds. Pressure can inherit substantial information from low-order normal momentum moments; signed shear can remain non-identifying because the incident normal--tangential correlation is lost by full-range averaging. This principle applies directly to kinetic--continuum coupling, Grad-type wall boundary conditions, and learned rarefied wall closures. The next step is not a larger generic neural network, but a reduced half-range state whose directional moments are selected from the wall functional and whose finite-distance transport to the surface is explicitly verified.

Three extensions follow naturally from these results and are within reach of the present pipeline. A gas-surface kernel sweep, replacing the fully diffuse wall by Maxwell or Cercignani--Lampis reflection with partial accommodation, would test how much of the sufficiency statement is carried by the incident state itself and how much by exact knowledge of the kernel. A Mach sweep of the collision pilot would separate compression-layer strength from rarefaction within the observability boundary. And the practical payoff, compact reduced representations of the incident half-range, whether directional flux quadratures or learned encodings sampled a mean free path from the surface, can now be developed against a benchmark whose answer is known, because $S_{\mathrm{HR}}$ bounds what any of them can achieve.

\section*{ACKNOWLEDGMENTS}
No external funding was received for this work. During the preparation of this manuscript the authors used ChatGPT based on GPT-5.6 Sol and Claude Fable 5 (Anthropic) together to improve language, readability, code review, and figure formatting; the authors reviewed and edited all generated material and take full responsibility for the content. The tools did not generate or alter the direct simulation Monte Carlo data, physical models, numerical results, or scientific conclusions.

\section*{AUTHOR DECLARATIONS}
\subsection*{Conflict of Interest}
The authors have no conflicts to disclose.
\subsection*{Author Contributions}
Elyas Lekzian: Conceptualization, DSMC data curation, geometry verification, rarefied-flow methodology and interpretation, validation, writing - review and editing. Ehsan Roohi: Conceptualization, machine-learning methodology, software, analysis, visualization, writing - original draft, writing - review and editing, supervision, corresponding author.

\section*{DATA AVAILABILITY}
Here SHA-256 denotes the 256-bit Secure Hash Algorithm, and JSON denotes JavaScript Object Notation.
The raw DSMC fields that support the 57-condition surrogate study are available from the corresponding author upon reasonable request. The reproducibility package contains machine-readable manifests for the 69 raw files and 57 deduplicated physical conditions. The available 51-case execution metadata are supplied with processed validation tables and SHA-256 checksums. The SPARTA packages contain block-resolved field moments, wall impulses and collision-tally metadata. For Table~\ref{tab:momentgate}, the package supplies the executable feature-extraction and ExtraTrees training scripts, all case-by-seed metrics, and the aggregate JSON record. It also includes the compact four-case $S_{\mathrm{off}}$ metrics used in Table~\ref{tab:offwallpilot}, together with the endpoint/intermediate split and normalized sampling offsets. Packed numerical arrays are supplied with validation reports and exact run inputs. The forward-facing precision extension contains ten 5000-step collision blocks for each intermediate Knudsen number. The repository provides the analysis scripts and source code for every figure.

\appendix
\section{Output-specialized surrogates and database-level validation}
\label{app:surrogate}
This appendix documents the surrogate infrastructure that supports the kinetic study in the main text: the design of the output-specialized coordinate-conditioned models, their blind validation on the 57-condition database, the direct wall-load emulation, the thermodynamic-consistency diagnostic, and the ablation record behind the design decisions.

\subsection{Model design and error metrics}
Here SiLU denotes the sigmoid linear unit activation, and AdamW denotes Adam optimization with decoupled weight decay.
The coordinate-conditioned surrogate used for the 57-case field study is still designed around the physical character of the outputs rather than around code convenience. A single network trained on all fields and all wall quantities with one loss is statistically attractive but physically mismatched: bulk temperature and velocity occupy the entire gas region and vary relatively smoothly, pressure is localized and positive in the compression layer, and wall coefficients are one-dimensional functions with sharp windward-face and apex peaks.

The smooth-field surrogate predicts $u$, $v$, and $T$ at gas-cell locations. Its case input contains Mach number, logarithmic Knudsen number, height ratio, wall-temperature ratio, and orientation. Its coordinate input contains the normalized coordinates $(\xi,\eta)$, smooth distance features to the wall, edges, and apex, and Fourier coordinate features. Hard region indicators were avoided because they produce block-like discontinuities in the reconstructed contours. The pressure-focused surrogate predicts $\log P$ and is trained with weights that emphasize high-pressure, near-wall, and near-apex cells. Predicting the logarithm stabilizes the positive pressure field and prevents a broad low-pressure region from dominating the loss. The wall-surface surrogate predicts $C_p$, $C_q$, and $|\tau|$ on the protrusion wall using the surface coordinate, face identity, tangent and normal direction, local midpoint, and apex distance. Here $C_p$ denotes the wall pressure coefficient, $C_q$ the wall heat-transfer coefficient used in the DSMC database, and $|\tau|$ the magnitude of the wall shear-stress coefficient. Separating the wall-surface surrogate prevents sparse wall data from being overwhelmed by hundreds of thousands of gas-cell samples.

The smooth-field and pressure surrogates use multiplicative branch--trunk multilayer perceptrons: case and coordinate features are encoded separately, multiplied in a latent space, and decoded through fully connected layers with SiLU activations. The smooth-field model uses 160 latent channels, four hidden layers of width 224, dropout 0.03, and AdamW with initial learning rate $7\times10^{-4}$. The pressure-focused model uses 192 latent channels, five hidden layers of width 256, dropout 0.01, and AdamW with initial learning rate $4\times10^{-4}$. The wall-surface surrogate uses 128 latent channels, hidden width 192, depth 3, dropout 0.03, AdamW with initial learning rate $8\times10^{-4}$, and peak/apex weighting to preserve localized wall extrema. All models use weight decay, Huber-type losses, gradient clipping, plateau-based learning-rate reduction, and early stopping. The reported widths, depths, latent dimensions, dropout rates, and initial learning rates were selected from compact pilot sweeps that compared validation relative-$L_2$ error, stability across epochs, and preservation of localized pressure and wall-load peaks. Smaller models under-resolved the apex and compression structures, whereas further increases in capacity produced negligible validation improvement and greater run-to-run variability. The selected settings were then fixed for all reported holdouts rather than retuned case by case. The training scripts, random seeds, processed split tables, and post-processing code are included with the submission package.

The main field metric is the relative $L_2$ error,
\begin{equation}
 \relLtwo(\phi)=\frac{\|\phi_{\rm pred}-\phi_{\rm DSMC}\|_2}{\|\phi_{\rm DSMC}\|_2},
\end{equation}
where ``pred'' denotes the neural prediction. For signed or near-zero diagnostics such as entropy-index error, a robust range-normalized percentage is used. If $S$ is the entropy-index field, the plotted error is
\begin{equation}
 E_S^{\%}=100\frac{S_{\rm pred}-S_{\rm DSMC}}{S_{{\rm DSMC},99}-S_{{\rm DSMC},1}},
\end{equation}
where $S_{{\rm DSMC},99}$ and $S_{{\rm DSMC},1}$ are the 99th and 1st percentiles of the DSMC entropy-index field for the same case. This definition avoids singular relative errors where the entropy proxy is close to zero.

\subsection{Validation performance}
Table~\ref{tab:validation} summarizes the main validation results. Temperature and streamwise velocity are the most accurate bulk variables because their dominant structures occupy a broad region of the flow. The transverse velocity is harder because it is smaller in magnitude and controlled by local deflection, weak recirculation, and recovery around the protrusion. Pressure remains more demanding than temperature because the high-pressure region is localized near the windward face and apex. The pressure-focused surrogate nevertheless improves the pressure prediction relative to the multi-output field model. Surface quantities are predicted with split-mean errors of 4.20--5.00\% in Table~\ref{tab:validation}, despite being sharply localized and sampled on a much smaller one-dimensional manifold. Independent same-condition DSMC realizations were not available, so a casewise particle-sampling confidence interval cannot be subtracted from these surrogate errors. The original grid, time-step, particle-per-cell, and sampling checks establish numerical adequacy of the reference database, while the remaining stochastic scatter is retained in the DSMC fields and wall profiles. Accordingly, differences of only a few percent are interpreted as error relative to the sampled DSMC mean rather than as statistically resolved departures from a separately measured DSMC noise floor.

\begin{table}[t]
\centering
\caption{Mean validation relative $L_2$ errors in percent for supported hold-out groups.}
\label{tab:validation}
\footnotesize
\begin{tabular}{llccc}
\toprule
Surrogate & Output & \shortstack{Mach--rarefaction--\\orientation} & \shortstack{Height\\interpolation} & \shortstack{Wall-temperature\\interpolation} \\
 & & $\Ma=6,\;\Kn=0.33$ & $h_p/h_s=1.25$ & $T_w/T_\infty=2$ \\
\midrule
Smooth field & $T$ & 3.28 & 2.23 & 2.00 \\
Smooth field & $u$ & 2.17 & 1.34 & 1.44 \\
Smooth field & $v$ & 4.80 & 4.19 & 3.39 \\
Pressure-focused & $P$ & 6.37 & 4.80 & 3.11 \\
Surface-only & $C_p,C_q,|\tau|$ mean & 5.00 & 4.20 & 4.42 \\
\bottomrule
\end{tabular}
\end{table}

Figure~\ref{fig:T1} uses a common three-column layout in each row: the left panel is the DSMC temperature field, the center panel is the surrogate prediction, and the right panel is the signed error normalized by the robust DSMC range. The horizontal and vertical axes are the normalized coordinates $\xi=(x-x_{\rm apex})/h_s$ and $\eta=(y-y_b)/h_s$, and the white triangle marks the solid protrusion. The comparison therefore shows both the location of the thermal structure and any displacement introduced by the surrogate. The physics behind the target field is worth stating because it explains where the surrogate succeeds and where it struggles. At Mach number 6 the protrusion forces the incoming stream to turn, and directed kinetic energy is converted into thermal energy in an oblique compression layer that leans over the windward face. In the transition regime this layer is not a thin discontinuity. Its thickness is set by a few local mean free paths, so at $\Kn=0.33$ the heated region is smeared over a distance comparable to the protrusion height, and increasing rarefaction thickens and weakens the layer while moving its temperature maximum away from the surface. Downstream of the apex, the expansion around the corner and the low-density leeward region produce a cooler recovery zone in which molecules re-equilibrate with the isothermal wall over many collision times. Across the three holdouts, the model reproduces this oblique thermal plume and its rarefaction-dependent diffuseness. For the Mach--rarefaction--orientation holdout, Table~\ref{tab:validation} gives a temperature relative-$L_2$ error of 3.28\%, and the corresponding contour captures the compression-heated region; the height-interpolation case recovers the altered FWD thermal footprint, in which the steeper effective obstruction strengthens and lifts the heated layer; and the wall-temperature case preserves the elevated near-wall band and downstream relaxation that develop when the wall is heated to twice the free-stream temperature. Residual error is concentrated near the apex and adjacent compression layer, where the gradients are steepest, and it is precisely there that small spatial shifts, cut-cell geometry, and DSMC scatter produce the largest pointwise differences. The error maps show no coherent bias over the smooth outer flow, which indicates that the coordinate-conditioned representation is not trading far-field accuracy for near-field structure.

Figures~\ref{fig:uval} and \ref{fig:vval} each contain three holdout rows and the same DSMC--prediction--error columns used for temperature; all panels are plotted in $\xi$--$\eta$ coordinates. Figure~\ref{fig:uval} evaluates the streamwise velocity, so its first two columns expose the broad acceleration and deceleration structure and its third column identifies positional error. The model captures the near-wall deceleration pocket that the compression layer imposes upstream of the obstacle, the acceleration of the deflected stream over the apex, and the outer high-speed region that remains close to the free-stream value. Two rarefaction effects are visible in the DSMC columns and are preserved by the surrogate. First, the gas does not come to rest on the surface: velocity slip of the order of a mean-free-path-scaled fraction of the local tangential velocity persists on both faces, so the near-wall band retains a finite speed instead of the no-slip zero of a continuum solution. The coordinate-conditioned features admit this behavior naturally because no boundary condition is hard-wired into the network. Second, the wake deficit behind the protrusion fills in over a distance that grows with Knudsen number, since fewer intermolecular collisions are available to diffuse momentum back toward the centerline of the disturbed region. The largest residuals lie along the near-wall shear layer and apex, where a small displacement of the low-speed region creates a large range-normalized difference. Figure~\ref{fig:vval} applies the same layout to transverse velocity, a smaller deflection-sensitive component. Physically, $v$ changes sign across the protrusion: the windward face pushes the stream upward, the leeward expansion turns it back toward the plate, and at the lower Knudsen numbers a weak recirculating region with negative near-wall $v$ appears in the orientation-dependent locations documented in the source DSMC study. Because the magnitude of $v$ is roughly one order below that of $u$, the same absolute positional uncertainty in the shear layer produces a visibly larger relative error, which is why the transverse component carries the largest field errors in Table~\ref{tab:validation} even though its sign and structure are recovered.

Pressure is the most localized bulk output and is treated separately. Figure~\ref{fig:pval} again uses three holdout rows; within each row, the left panel is DSMC pressure, the center panel is the pressure-focused prediction, and the right panel is robust-range-normalized signed error in $\xi$--$\eta$ coordinates. The layout makes the windward compression maximum, its downstream decay, and any spatial shift of the peak directly comparable. The surrogate captures the high-pressure region near the windward face and its downstream decay. The physical origin of the difficulty is that pressure, unlike temperature, is set by the normal momentum flux of a population that is strongly out of equilibrium inside the compression layer: the incoming high-speed molecules and the slow, hot, reflected molecules coexist in the same cells, and the resulting pressure maximum is confined to a thin band hugging the windward face and apex. Increasing the Knudsen number lowers the peak because fewer collisions are available to convert the directed free-stream flux into random thermal motion near the surface, so the compression signature spreads and weakens; the surrogate must therefore represent simultaneously a sharp near-wall band at $\Kn=0.1$ and a diffuse one at $\Kn=0.8$. The BWD holdout has the largest pressure error because pressure depends nonlinearly on the combined Mach number, Knudsen number, and orientation, whereas the supported height and wall-temperature directions are resolved more accurately. This behavior is consistent with pressure peaks being controlled by orientation-dependent compression and molecular momentum transfer, and it motivated the logarithmic target and the peak-weighted loss described above. Figures~\ref{fig:T1}, \ref{fig:uval}, \ref{fig:vval}, and \ref{fig:pval} follow on consecutive pages.

A brief justification of the modeling choices closes this part. Randomized tree ensembles carry the equal-capacity comparison of the main text because that test needs a learner whose capacity is trivial to hold fixed while the input state changes; trees have no architecture to retune per state, are insensitive to monotone feature rescalings, and reproduce exactly from fixed seeds. The neural surrogates documented here carry the opposite duty: they must represent smooth fields with sharp near-wall structure across a parameter family, and there the coordinate conditioning, the solid-distance features, and the output specialization earn their complexity. Neither family is presented as optimal. Both are instruments, and the design target throughout was that no headline conclusion of the paper should move if either instrument were replaced by a reasonable alternative, a property the ablation record at the end of this appendix documents.

\clearpage
\begin{widepage}
\begin{figure}[H]
\centering
\includegraphics[width=0.80\linewidth]{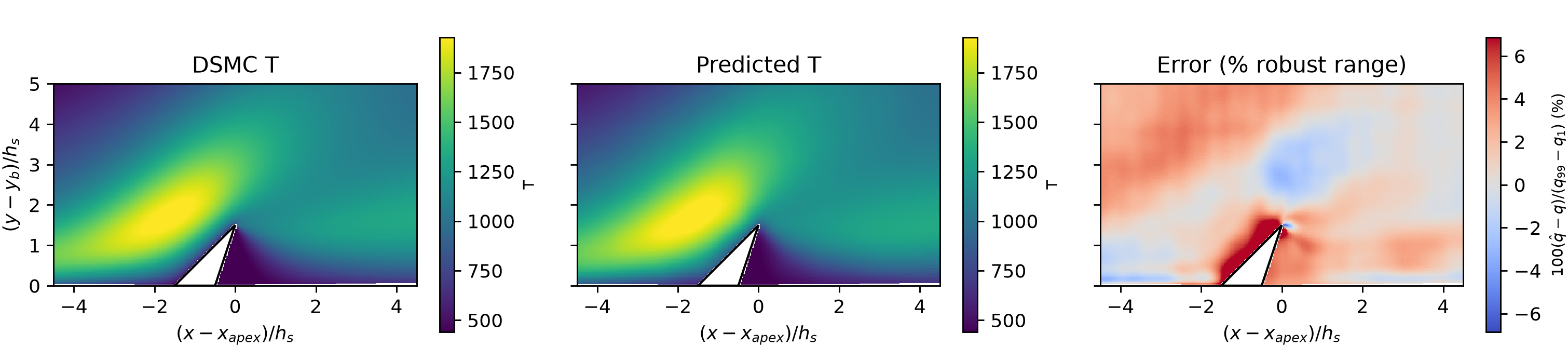}\\[-1mm]
\includegraphics[width=0.80\linewidth]{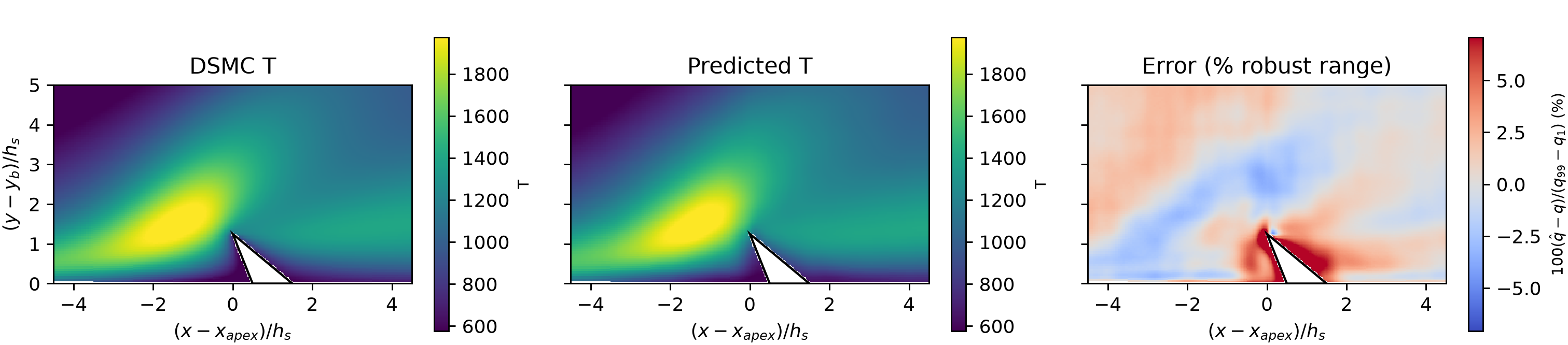}\\[-1mm]
\includegraphics[width=0.80\linewidth]{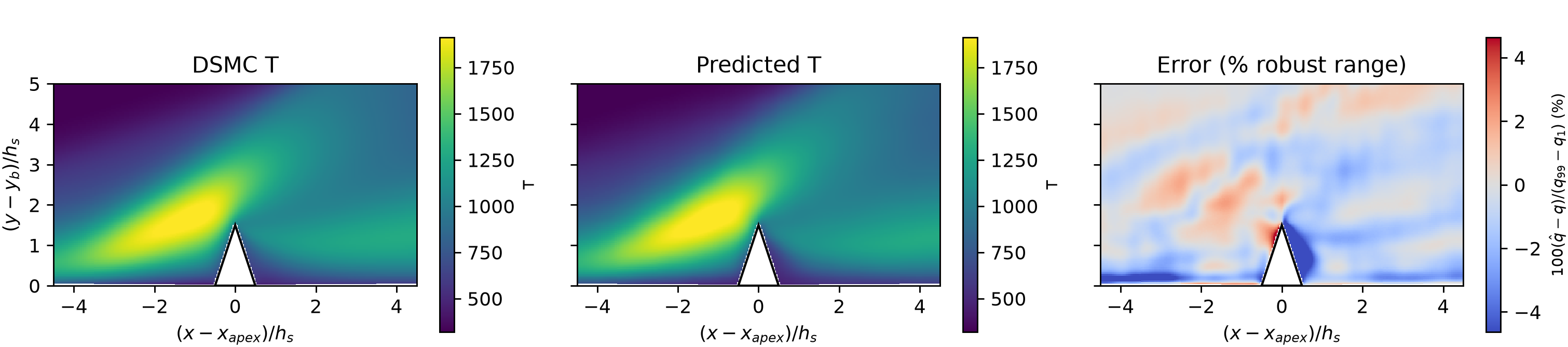}
\caption{Temperature validation for three representative holdouts. From top to bottom: Mach--rarefaction--orientation interpolation for BWD at $\Ma=6$, $\Kn=0.33$, $h_p/h_s=1.5$, and $T_w/T_\infty=1$; height interpolation for FWD at $h_p/h_s=1.25$; and wall-temperature interpolation for ISO at $\Kn=0.1$ and $T_w/T_\infty=2$. Each row shows DSMC, surrogate prediction, and robust-range-normalized error. Temperature color scales are in kelvin. The oblique compression-heated plume, its rarefaction-dependent thickness, and the cooler leeward recovery region are reproduced; residual error concentrates at the apex where gradients are steepest.}
\label{fig:T1}
\end{figure}
\end{widepage}

\clearpage
\begin{widepage}
\begin{figure}[H]
\centering
\includegraphics[width=0.80\linewidth]{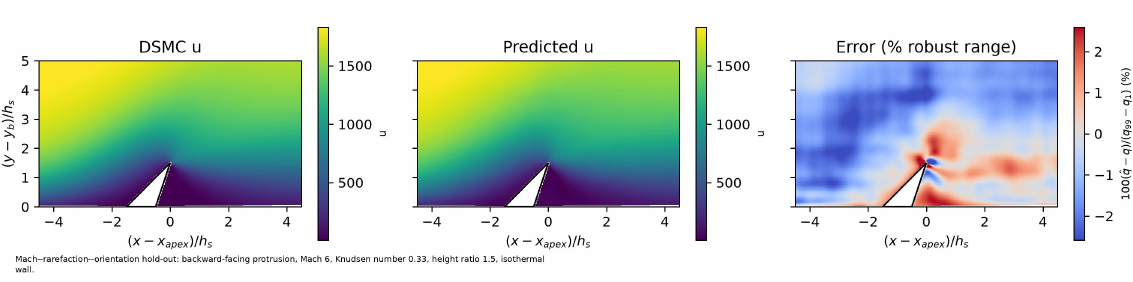}\\[-1mm]
\includegraphics[width=0.80\linewidth]{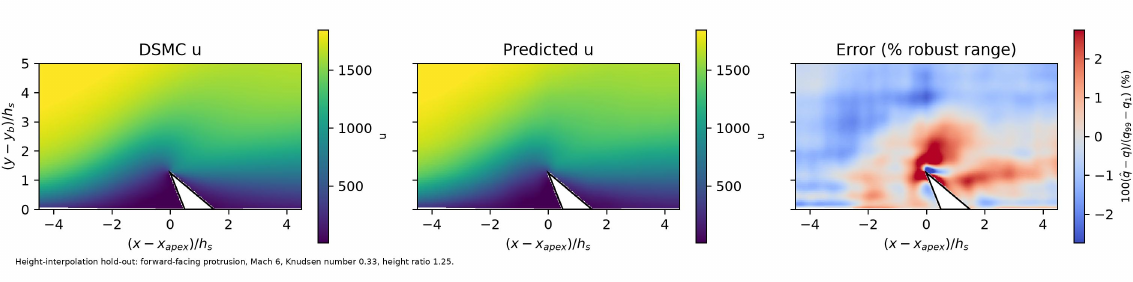}\\[-1mm]
\includegraphics[width=0.80\linewidth]{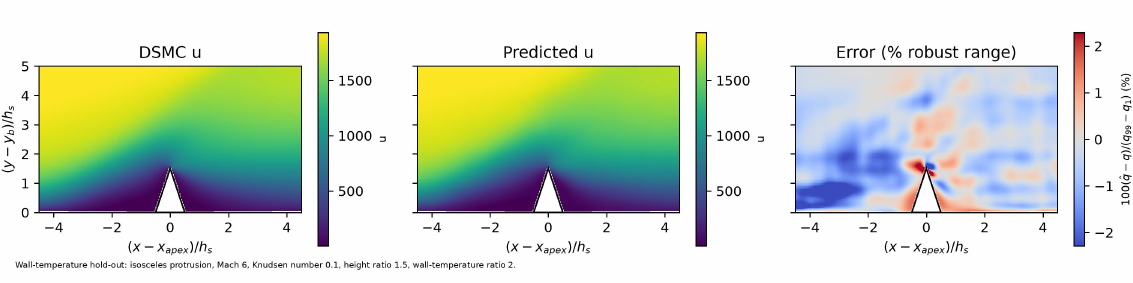}
\caption{Streamwise-velocity validation for the three representative holdout cases. Row captions below each panel identify the physical validation condition. Velocity color scales are in $\mathrm{m\,s^{-1}}$. The model captures the upstream deceleration pocket, the accelerated deflected stream above the apex, the finite slip velocity retained at the wall, and the slow wake recovery, while errors remain concentrated near the wall and apex.}
\label{fig:uval}
\end{figure}
\end{widepage}

\clearpage
\begin{widepage}
\begin{figure}[H]
\centering
\includegraphics[width=0.80\linewidth]{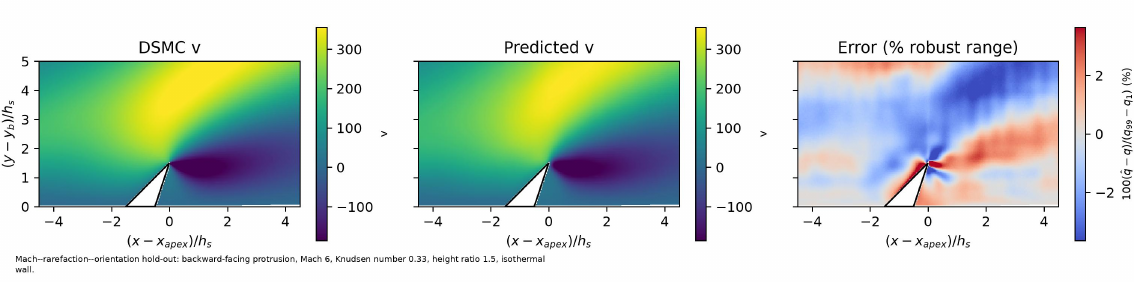}\\[-1mm]
\includegraphics[width=0.80\linewidth]{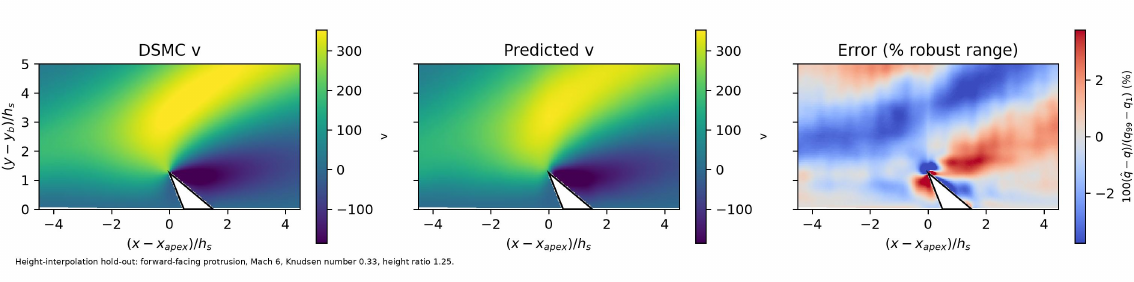}\\[-1mm]
\includegraphics[width=0.80\linewidth]{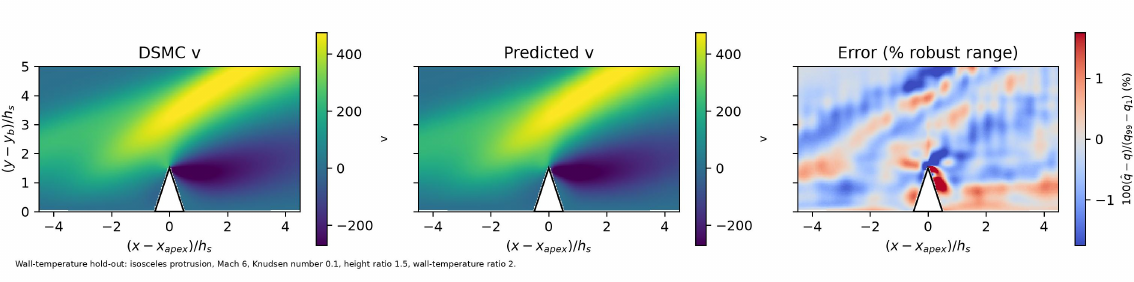}
\caption{Transverse-velocity validation for the same holdout cases. Velocity color scales are in $\mathrm{m\,s^{-1}}$. This component is physically tied to streamline deflection and weak recirculation: it is positive over the windward face, changes sign in the leeward expansion, and is an order of magnitude smaller than $u$, so it is more sensitive to small positional errors in the shear layer and recovery region.}
\label{fig:vval}
\end{figure}
\end{widepage}

\clearpage
\begin{widepage}
\begin{figure}[H]
\centering
\includegraphics[width=0.80\linewidth]{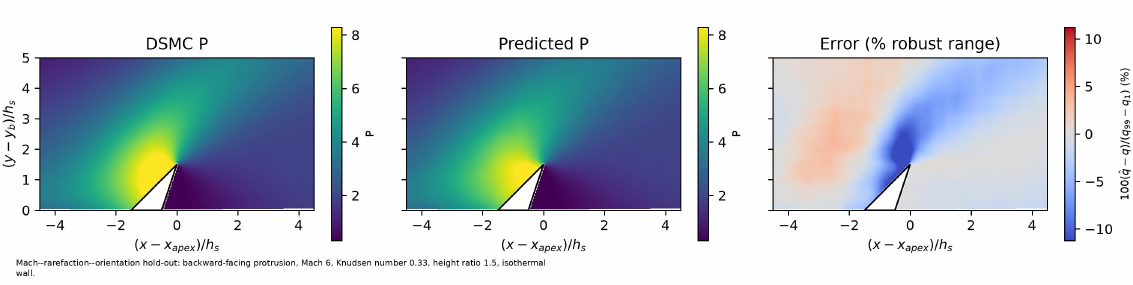}\\[-1mm]
\includegraphics[width=0.80\linewidth]{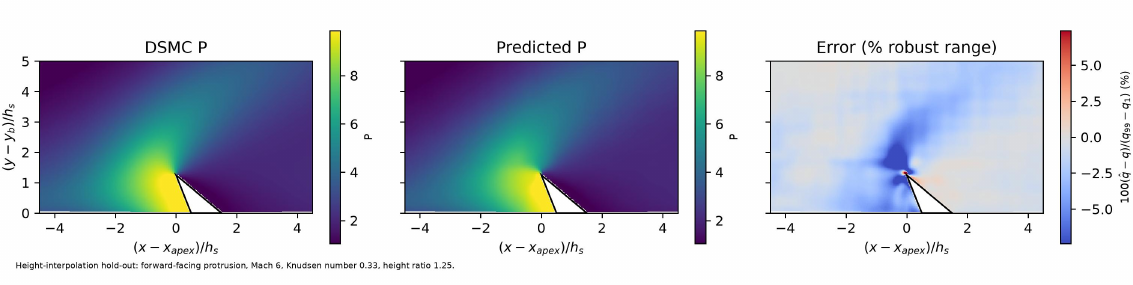}\\[-1mm]
\includegraphics[width=0.80\linewidth]{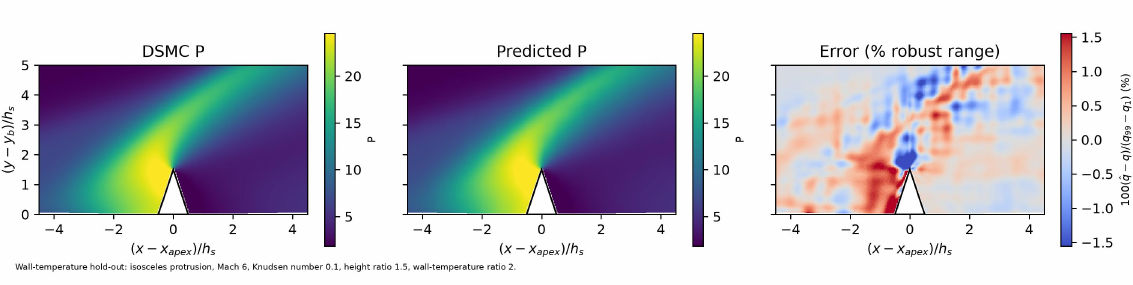}
\caption{Pressure validation using the pressure-focused surrogate. Pressure color scales are in pascals. The pressure model is trained on logarithmic pressure with emphasis on high-pressure, near-wall, and apex-adjacent cells. This specialization improves the peak-compression region relative to a single multi-output field network; the remaining error sits in the thin windward band where the incident and reflected molecular populations coexist far from equilibrium.}
\label{fig:pval}
\end{figure}
\end{widepage}

\subsection{Surface-load prediction}
The protrusion-wall response is the most direct engineering output of the surrogate because structural load and thermal protection depend on the wall pressure, heat transfer, and shear distribution. Figure~\ref{fig:surface} is organized by validation group in rows and by wall quantity in columns. In every panel, the horizontal axis is the normalized protrusion coordinate running from one base endpoint through the apex near the face transition to the opposite endpoint; the vertical axis is $C_p$, $C_q$, or $|\tau|$, and DSMC and surrogate profiles are superposed. This representation tests the complete wall distribution rather than only a peak value. The windward face produces a monotone rise in $C_p$ and $C_q$ toward the apex, followed by a sharp drop on the leeward side, and the physical mechanism behind each curve differs in an instructive way. The pressure coefficient grows along the windward face because the local surface inclination to the oncoming stream increases the incident normal momentum flux, and molecules arriving there have experienced progressively less shielding by the compression layer as the apex is approached. The heat-transfer coefficient follows the same rise because the same incident molecules carry the free-stream kinetic energy that the cold wall must accommodate; its peak sits at or just below the apex, where compression heating of the gas adjacent to the wall is strongest. On the leeward face both coefficients collapse: the expansion around the apex rarefies the local gas, part of the face lies in the ballistic shadow of the protrusion at the higher Knudsen numbers, and the few arriving molecules transfer little momentum or energy. The shear magnitude is the noisiest of the three wall functions because it is the small tangential residual of two nearly canceling contributions, the tangential momentum delivered by incident molecules and the tangential momentum removed by diffusely re-emitted ones, so its statistical convergence is intrinsically slower at fixed sample size. The face-aware wall-surface surrogate captures this asymmetry and preserves the peak location. No post-hoc smoothing is applied to the DSMC wall profiles: the visible point-to-point fluctuations, particularly in $|\tau|$, are retained because smoothing could alter the peak magnitude and location that the surrogate is intended to reproduce. Remaining discrepancies are concentrated near the apex and face transition, where the response changes over a short arc length and DSMC statistics are most sensitive to cut-cell geometry and particle sampling. Table~\ref{tab:surface} reports the split-averaged surface errors.

The heat-transfer coefficient deserves one added remark because it has no counterpart in the kinetic part of the paper. $C_q$ is comparatively forgiving for a field-trained surrogate: its windward rise is governed by the same incident energy flux that shapes the thermal plume the model already represents, its leeward values are small, and unlike shear it carries no sign change to protect, so it can be recovered from the same near-wall structure that encodes the temperature field. This is the emulator-level shadow of the hierarchy that the kinetic analysis of the main text makes precise: wall quantities tied to the incident normal energy and momentum fluxes are encoded far more robustly in nearby fields than the small signed tangential residual is. The face-resolved display also matters here. Split-averaged scalars can hide compensating errors between windward and leeward faces, and the row layout of the surface figure was chosen so that any such compensation would be visible directly rather than absorbed into a mean.

One more distinction between the two windward peaks is worth recording. The pressure maximum sits essentially at the apex, where the incident normal momentum flux is largest, while the heat-transfer maximum can sit marginally below it, because energy accommodation weighs the full kinetic energy of the arriving molecules rather than only their normal momentum, and the hottest arriving population has crossed the thickest part of the compression layer. The surrogate preserves the ordering and location of the two peaks, which is a stricter test of near-wall fidelity than either peak value alone.

\clearpage
\begin{widepage}
\begin{figure}[H]
\centering
\includegraphics[width=0.78\linewidth]{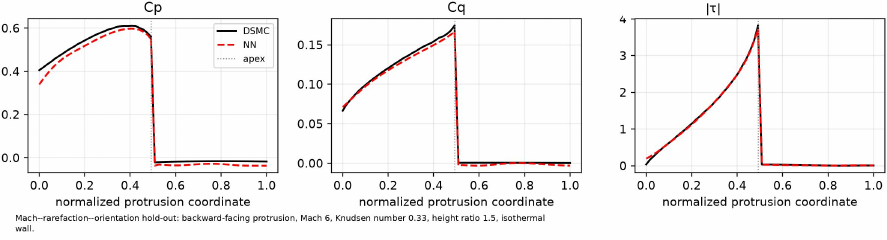}\\[-1mm]
\includegraphics[width=0.78\linewidth]{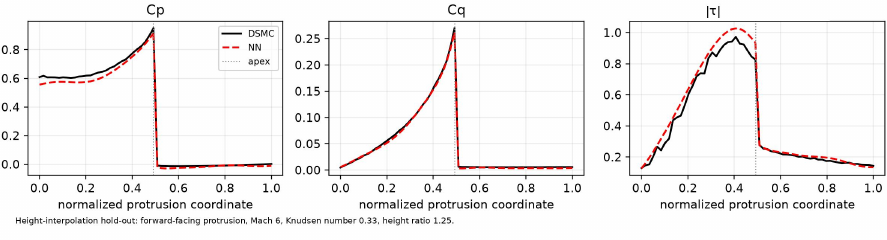}\\[-1mm]
\includegraphics[width=0.78\linewidth]{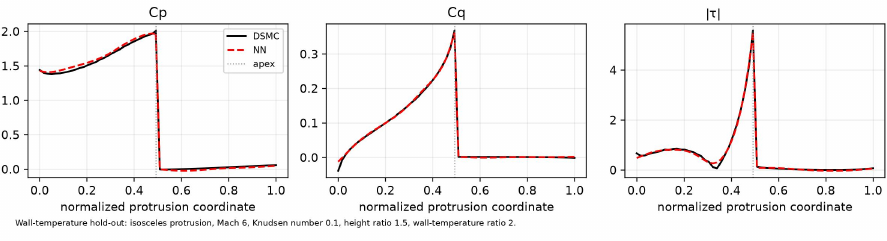}
\caption{Surface-only validation of protrusion-wall $C_p$, $C_q$, and $|\tau|$. DSMC curves are shown without post-hoc smoothing so that particle-sampling fluctuations and localized extrema remain visible. The wall-surface surrogate is trained on face-aware wall features rather than on the bulk field. The agreement is strongest on the windward rise and peak region, which are the most important parts of the protrusion-wall response for pressure and thermal loading.}
\label{fig:surface}
\end{figure}
\end{widepage}

\begin{table}[t]
\centering
\caption{Surface validation summary. Values are relative $L_2$ errors in percent.}
\label{tab:surface}
\begin{tabular}{lccc}
\toprule
Validation group & $C_p$ & $C_q$ & $|\tau|$ \\
\midrule
Mach--rarefaction--orientation holdout & 5.23 & 3.32 & 6.45 \\
Height-interpolation holdout & 3.83 & 3.12 & 5.65 \\
Wall-temperature-interpolation holdout & 3.90 & 4.37 & 4.97 \\
\bottomrule
\end{tabular}
\end{table}

\subsection{Physics-aware post-processing}
Field contours alone are not sufficient for evaluating a surrogate for rarefied hypersonic protrusion flow. A visually smooth field can still miss wall loads, peak locations, or thermodynamic consistency. We therefore compute engineering quantities of interest (QoIs), region-conditioned errors, and an entropy-index diagnostic. The 57-condition field data used in this section contains macroscopic variables but not the velocity distributions exported by the new kinetic pilot. Consequently, we do not claim to compute Boltzmann entropy production from the data. Instead, for a monatomic gas we define a local-equilibrium entropy-rise proxy,
\begin{equation}
 \frac{\Delta s_{\rm eq}}{R}=\frac{\gamma}{\gamma-1}\ln\left(\frac{T}{T_{\rm ref}}\right)-\ln\left(\frac{P}{P_{\rm ref}}\right), \qquad \gamma=\frac{5}{3},
\end{equation}
where $R$ is the specific gas constant and $T_{\rm ref}$ and $P_{\rm ref}$ are estimated from an upstream/outer DSMC region for each case. This diagnostic uses the temperature from the smooth-field surrogate and the pressure from the pressure-focused surrogate, so it checks whether two separately trained decoders remain thermodynamically compatible.

Figures~\ref{fig:entropy1} and \ref{fig:entropy2} use rows for the three protrusion orientations and columns for the DSMC entropy index, the index reconstructed from the independent temperature and pressure surrogates, and their robust-range-normalized percent difference. The horizontal and vertical axes are $\xi$ and $\eta$, and the triangular solid is masked. This layout tests both the spatial entropy-rise pattern and compatibility of the separately trained decoders. The DSMC and reconstructed maps contain the same high-entropy band associated with compression and thermalization, while the largest residuals occur near the apex and near-wall recovery region where temperature and pressure gradients are steep. Table~\ref{tab:entropy} shows that the range-scaled mean absolute error remains below 3.3\% for every validation group.

\clearpage
\begin{widepage}
\begin{figure}[H]
\centering
\includegraphics[width=0.78\linewidth]{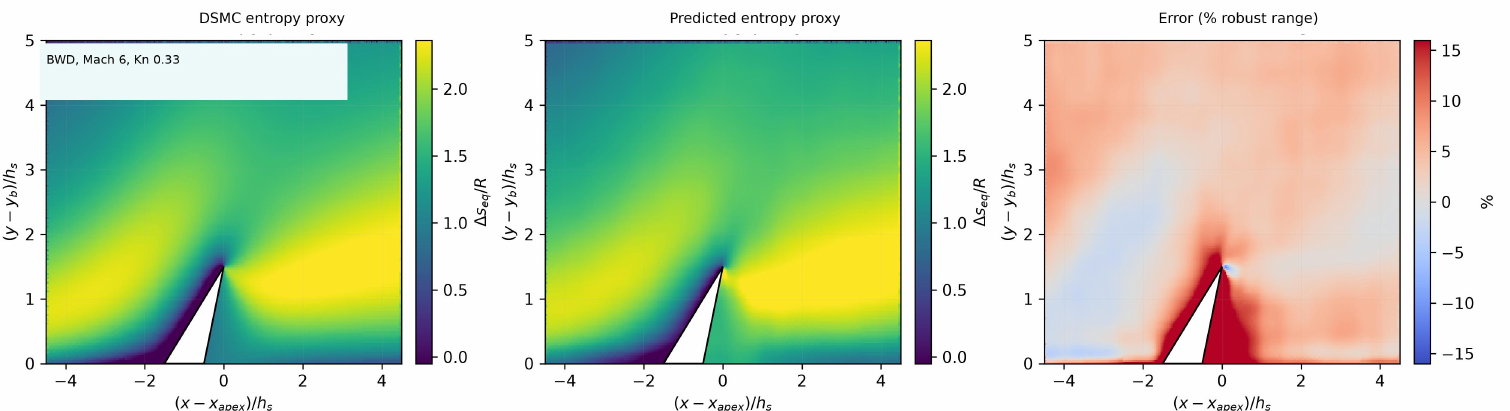}\\[-1mm]
\includegraphics[width=0.78\linewidth]{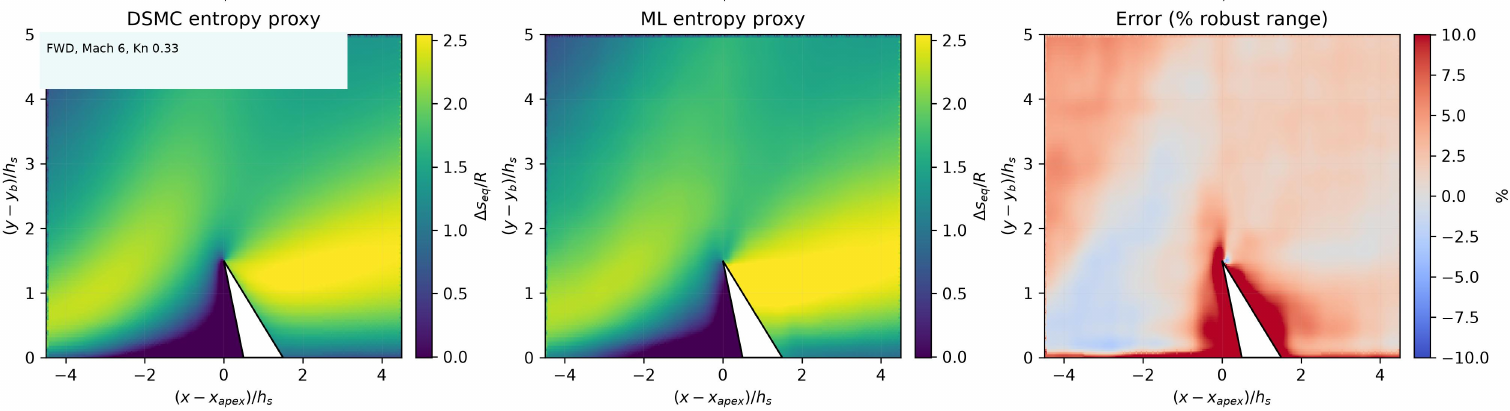}\\[-1mm]
\includegraphics[width=0.78\linewidth]{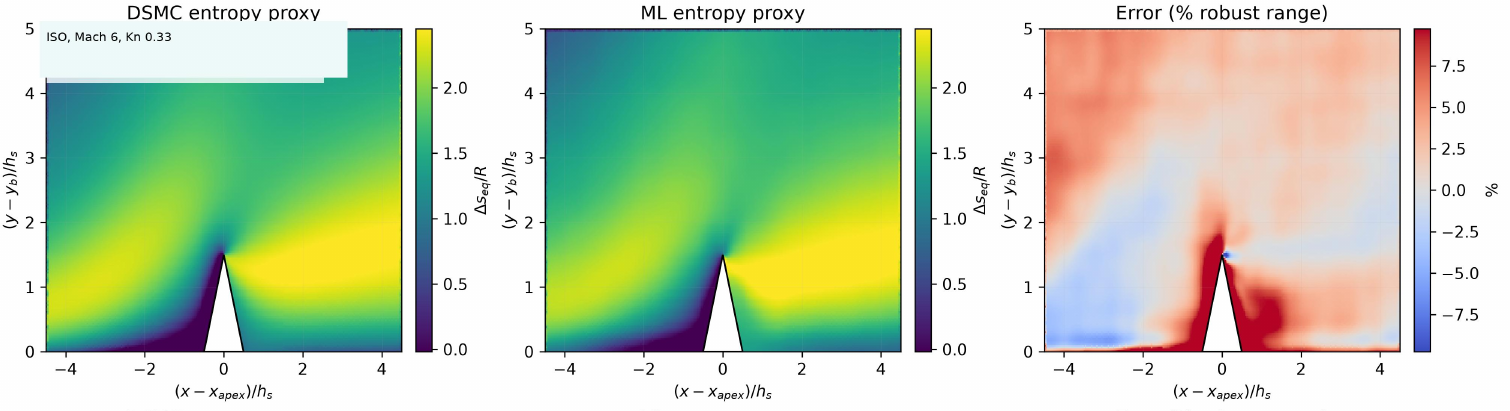}
\caption{Local-equilibrium entropy-rise proxy for the Mach--rarefaction--orientation validation cases. The solid triangle is masked, and the error is reported as a percent of the robust DSMC entropy-index range. The main entropy-rise structure is reproduced, while the residual error is concentrated around the apex.}
\label{fig:entropy1}
\end{figure}
\end{widepage}

\clearpage
\begin{widepage}
\begin{figure}[H]
\centering
\includegraphics[width=0.78\linewidth]{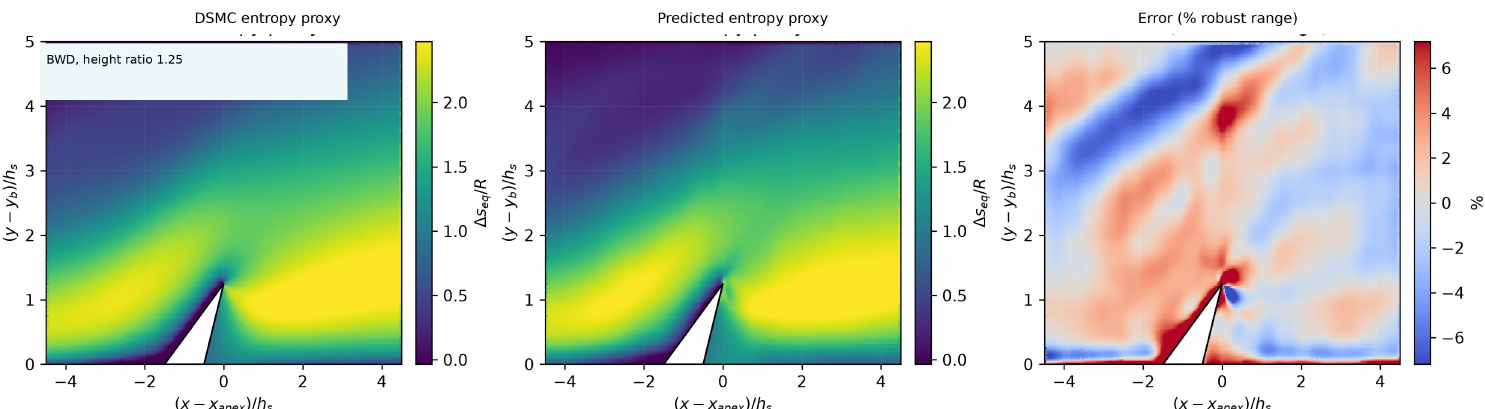}\\[-1mm]
\includegraphics[width=0.78\linewidth]{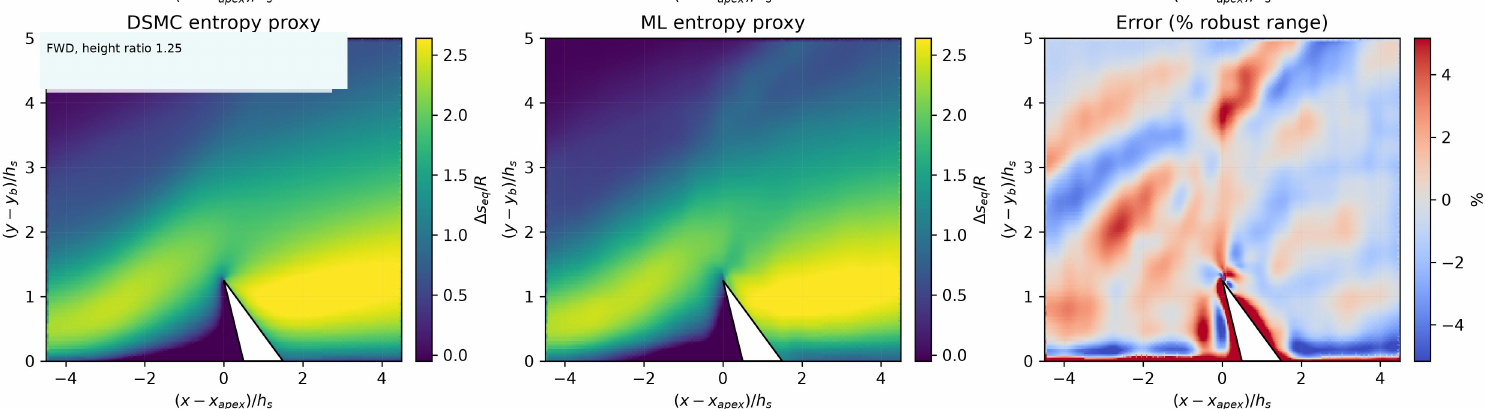}\\[-1mm]
\includegraphics[width=0.78\linewidth]{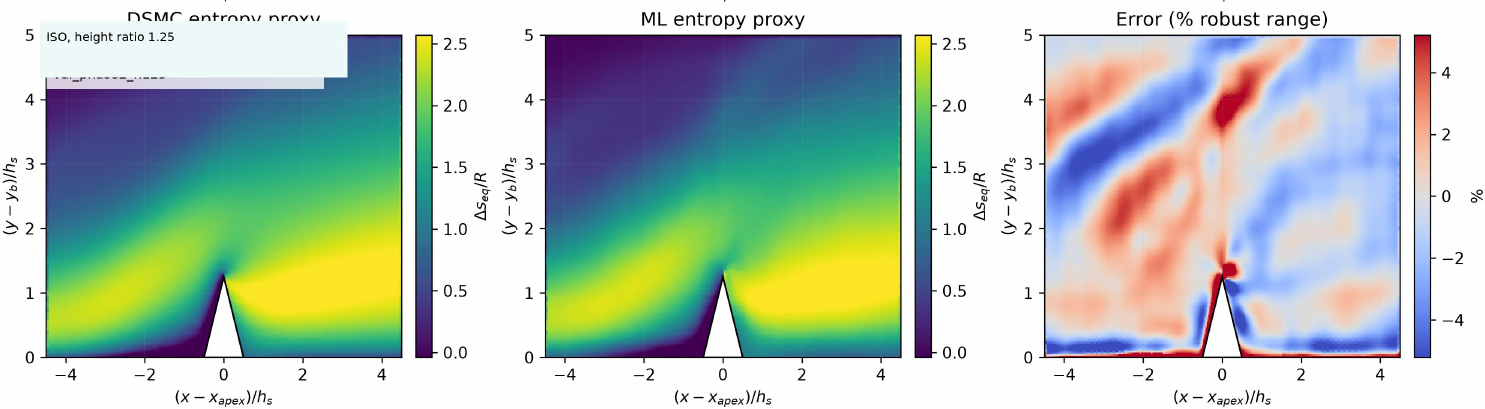}
\caption{Local-equilibrium entropy-rise proxy for height-interpolation validation cases. The comparison shows that the separately trained temperature and pressure surrogates remain mutually consistent when combined in a thermodynamic post-processing diagnostic.}
\label{fig:entropy2}
\end{figure}
\end{widepage}

The physical content of the entropy check is easy to state. The proxy combines temperature and density into the ideal-gas entropy index, so it is a nonlinear compound of two separately predicted fields, and a surrogate that satisfied each field marginally while violating their thermodynamic coupling would fail here even with small individual errors. The maps show the expected structure. Entropy rises across the thickened compression layer, where directed kinetic energy is irreversibly thermalized, peaks in the hot near-apex region, and relaxes through the leeward expansion, while the near-wall band reflects the isothermal boundary rather than the interior balance, which is why the residuals concentrate there. Passing this test does not certify Boltzmann-level thermodynamics and is not claimed to; it certifies that the emulated fields are mutually consistent at the level a continuum observer would check first.

\begin{table}[t]
\centering
\caption{Entropy-index and peak-location validation. $S$ denotes $\Delta s_{\rm eq}/R$; range mean absolute error is normalized by the robust DSMC entropy range; $d_{\rm peak}(P)$ is the normalized distance between predicted and DSMC pressure-peak locations.}
\label{tab:entropy}
\footnotesize
\begin{tabular}{lccccc}
\toprule
Validation group & \shortstack{$\relLtwo(S)$\\(\%)} & \shortstack{Range mean\\absolute error (\%)} & \shortstack{$|\Delta T_{\max}|$\\(K)} & $|\Delta P_{\max}|$ & \shortstack{$d_{\rm peak}(P)$\\normalized} \\
\midrule
Mach--rarefaction--orientation & 9.01 & 3.27 & 10.86 & 0.77 & 0.136 \\
Height interpolation & 6.77 & 2.10 & 20.17 & 1.38 & 0.158 \\
Wall-temperature interpolation & 5.43 & 1.82 & 8.71 & 0.64 & 0.113 \\
\bottomrule
\end{tabular}
\end{table}

Region-conditioned errors are reported in Table~\ref{tab:region}. The near-wall and near-apex zones are more difficult than the outer flow, especially for transverse velocity and entropy. This is physically expected because near-wall velocity and entropy depend on gas-surface interaction, cut-cell geometry, and local particle statistics. The compression-region pressure error remains larger than the global average, confirming that pressure is the most demanding field and justifying the separate pressure surrogate.

\begin{table}[t]
\centering
\caption{Selected region-conditioned relative $L_2$ errors in percent.}
\label{tab:region}
\begin{tabular}{llccccc}
\toprule
Validation group & Region & $u$ & $v$ & $T$ & $P$ & $S$ \\
\midrule
Mach--rarefaction--orientation & all & 1.39 & 5.30 & 3.58 & 6.35 & 9.01 \\
Mach--rarefaction--orientation & near apex & 4.14 & 7.94 & 5.96 & 8.25 & 16.74 \\
Mach--rarefaction--orientation & near wall & 8.57 & 21.11 & 6.62 & 3.98 & 21.25 \\
Height interpolation & all & 0.92 & 4.37 & 2.58 & 4.74 & 6.77 \\
Height interpolation & near apex & 4.72 & 8.18 & 4.17 & 6.61 & 11.49 \\
Wall-temperature interpolation & all & 0.99 & 3.73 & 2.28 & 3.36 & 5.43 \\
Wall-temperature interpolation & near wall & 8.45 & 21.83 & 5.45 & 2.83 & 13.11 \\
\bottomrule
\end{tabular}
\end{table}

The engineering QoI tests provide a compact way to evaluate whether the surrogate preserves peaks and integrated loads rather than only cell-wise error. Table~\ref{tab:qoi} reports parity coefficients for representative field, entropy, and surface quantities. Mean temperature, mean pressure, mean speed, peak temperature, and peak pressure are reproduced with high parity. The peak entropy-index statistic is excluded from Table~\ref{tab:qoi}: its low parity ($R^2=0.272$) shows that a single localized maximum is unstable under DSMC scatter even when the global entropy structure is reproduced. Surface QoIs show particularly strong parity because the face-aware wall-surface surrogate is trained directly on the wall profiles and preserves both the windward rise and apex peak.

\begin{table}[t]
\centering
\caption{Coefficient of determination between DSMC-derived and predicted engineering QoIs. Values close to one indicate strong parity.}
\label{tab:qoi}
\begin{tabular}{lcc@{\qquad}lcc}
\toprule
Field/entropy QoI & Symbol & $R^2$ & Surface QoI & Symbol & $R^2$ \\
\midrule
Peak temperature & $T_{\max}$ & 0.992 & Peak pressure coefficient & $C_{p,\max}$ & 0.999 \\
Peak pressure & $P_{\max}$ & 0.987 & Peak heat-transfer coefficient & $C_{q,\max}$ & 0.998 \\
Mean temperature & $\overline{T}$ & 0.997 & Peak shear magnitude & $|\tau|_{\max}$ & 1.000 \\
Mean pressure & $\overline{P}$ & 1.000 & Integrated pressure coefficient & $\int C_p\,ds$ & 0.999 \\
Mean speed & $\overline{|\mathbf{u}|}$ & 0.998 & Integrated heat-transfer coefficient & $\int C_q\,ds$ & 0.995 \\
Mean entropy index & $\overline{S}$ & 0.977 & Integrated shear magnitude & $\int |\tau|\,ds$ & 0.996 \\

\bottomrule
\end{tabular}
\end{table}

\subsection{Interpretation of the validation trends and design ablation}
The validation results are consistent with the DSMC physics of triangular protrusions. FWD protrusions are the most demanding orientation because the incoming flow meets a steeper windward face earlier, producing stronger compression and larger pressure and heat-transfer peaks. In a continuum interpretation this would be described as a strong obstacle-induced compression layer; in the rarefied setting the same effect is mediated by molecular trajectories and wall accommodation. The surrogate captures the dominant response but still shows its largest errors where the compression layer impinges on the apex, which is the correct physical location for difficulty.

Height changes the effective obstruction and the size of the separated or recirculating region. The reference DSMC study already characterized the associated orientation-dependent vortices: forward-facing cases concentrate recirculation on the windward side, backward-facing cases emphasize the leeward wake, and the symmetric protrusion develops a stronger recirculating region as height increases. We therefore use vortex topology as physical context for the load footprints rather than duplicating that established classification with another contour set. The height-sweep validation at $h_p/h_s=1.25$ lies inside the sampled height range and tests whether the model can interpolate the growth of the thermal and pressure footprints. The model performs well in this case because, over the sampled interval, the original DSMC study reports nearly monotone and approximately low-order sensitivity of the peak quantities to height. The model should not be interpreted as learning arbitrary height-Mach interaction, because such combinations were not sampled. Wall temperature affects energy accommodation and near-wall thermal exchange. Increasing wall temperature raises the local gas temperature while reducing the heat-transfer coefficient when the wall becomes closer to the gas temperature. The wall-temperature holdout at $T_w/T_\infty=2$ confirms that the surrogate captures this supported thermal direction, with low errors in temperature and pressure and moderate errors in heat-transfer coefficient.

Rarefaction changes the balance between collision-mediated transport and ballistic molecular motion. Increasing Knudsen number reduces pressure loading and drag in the DSMC database while changing the thickness and location of thermal structures. The Mach-rarefaction-orientation block is therefore the strongest and most stringent validation block, because it contains the full factorial interaction of Mach number, Knudsen number, and orientation. The pressure error is largest there because pressure responds nonlinearly to both compression and rarefaction. This observation is not a weakness of the framework; it explains why pressure was separated into its own specialized head.

The ablation summary in Table~\ref{tab:ablation} documents the design decisions. The most important result is that geometry correction is not optional. The wrong base interpretation creates artificial solid regions and corrupts both field and surface representations. The second result is that pressure and surface loads require dedicated decoders. These choices follow from the physics of localized compression and wall loading, rather than from arbitrary neural-network tuning.

\begin{table}[t]
\centering
\caption{Ablation and design-decision summary.}
\label{tab:ablation}
\footnotesize
\begin{tabular}{lll}
\toprule
\parbox[t]{0.22\textwidth}{Design choice} & \parbox[t]{0.50\textwidth}{Observed effect} & Decision \\
\midrule
\parbox[t]{0.22\textwidth}{Bounding-box extrema as base endpoints} & \parbox[t]{0.50\textwidth}{Creates artificial right-triangle geometry for forward- and backward-facing cases and contaminates masks, distance features, and contours.} & Rejected \\
\parbox[t]{0.22\textwidth}{Fixed base plus orientation-specific apex} & \parbox[t]{0.50\textwidth}{Recovers the DSMC geometry and removes nonphysical solid-region artifacts.} & Used \\
\parbox[t]{0.22\textwidth}{Single multi-output field model for pressure} & \parbox[t]{0.50\textwidth}{Under-resolves localized peak compression; the Mach--rarefaction--orientation pressure error remains about 8.82\%.} & Baseline \\
\parbox[t]{0.22\textwidth}{Pressure-focused logarithmic model} & \parbox[t]{0.50\textwidth}{Reduces the Mach--rarefaction--orientation pressure error to 6.37\% and improves the compression-region structure.} & Used \\
\parbox[t]{0.22\textwidth}{Unified field/surface model} & \parbox[t]{0.50\textwidth}{Smooths surface peaks because sparse wall samples are diluted by the much larger gas-cell data set.} & Rejected \\
\parbox[t]{0.22\textwidth}{Face-aware surface-only model} & \parbox[t]{0.50\textwidth}{Preserves windward/leeward asymmetry and predicts $C_p$, $C_q$, and $|\tau|$ at about 4--5\% mean error.} & Used \\
\parbox[t]{0.22\textwidth}{Finite full-range hierarchy $S_0$--$S_2$} & \parbox[t]{0.50\textwidth}{$P_{ij}$ improves blind $C_p$ relative to primitives, but no finite state improves transferable signed shear and none establishes uniqueness.} & Insufficient for traction closure \\
\parbox[t]{0.22\textwidth}{Incident half-range state $S_{\mathrm{HR}}$} & \parbox[t]{0.50\textwidth}{Recovers co-temporal $C_p$ within 0.07--0.28\% and signed $C_f$ within 1.30--7.35\%, with independent-window pressure confirmation.} & Sufficiency benchmark \\
\bottomrule
\end{tabular}
\end{table}

\section{DSMC execution metadata and reproducibility}
\label{app:dsmc_execution}
CPU denotes central processing unit, and $\tau_{\rm coll}$ denotes the mean collision time used to normalize the time step $\Delta t$.
Reproduction of the reference database requires the physical-condition keys together with the numerical scale of each particle calculation. The execution-accounting sheet supplied for this study contains 51 run records. For every recorded run it specifies Knudsen number, Mach number, protrusion orientation, height ratio $h_p/h_s$, wall-temperature ratio $T_w/T_\infty$, instantaneous particle population, CPU-core allocation, reported runtime, and normalized time step $\Delta t/\tau_{\rm coll}$. The record covers the complete 27-case Mach-number--Knudsen-number--orientation validation block, 12 height-sweep runs at $h_p/h_s=1.0$ and $1.25$, and 12 wall-temperature-sweep runs at $T_w/T_\infty=2$ and $4$. Table~\ref{tab:dsmc_execution} gives a compact summary; the exact case-level values are provided in \texttt{data/dsmc\_execution\_metadata\_51\_cases.csv}, together with the source accounting document and SHA-256 checksums.

Across the supplied records, the simulations contain approximately $1.46$--$2.00$ million particles, use either 14 or 26 CPU cores, report runtimes from 8.04 to 21.43~h, and maintain $\Delta t/\tau_{\rm coll}=0.16$--$0.19$. The complete physical-condition manifest contains six additional height-sweep cases at $h_p/h_s=2.0$. Their physical keys and data files are included in the 57-condition manifest, but corresponding execution-accounting entries were not present in the supplied run sheet; they are therefore excluded from the runtime ranges in Table~\ref{tab:dsmc_execution} rather than assigned inferred values.

\begin{table}[h]
\centering
\caption{Summary of the supplied DSMC execution records. Particle populations and reported runtimes are shown as minimum--median--maximum. The full case-level record is distributed with the reproducibility package.}
\label{tab:dsmc_execution}
\scriptsize
\setlength{\tabcolsep}{4pt}
\begin{tabular}{lccccc}
\toprule
Sampling block & Records & Particles ($10^6$) & CPU cores & Reported runtime (h) & $\Delta t/\tau_{\rm coll}$ \\
\midrule
Mach--rarefaction--orientation & 27 & 1.466--1.480--1.491 & 14, 26 & 8.05--8.70--18.15 & 0.16--0.19 \\
Height sweep (recorded subset) & 12 & 1.489--1.973--2.005 & 14, 26 & 8.63--11.07--21.43 & 0.17--0.19 \\
Wall-temperature sweep & 12 & 1.464--1.483--1.499 & 26 & 8.04--8.74--9.10 & 0.17--0.19 \\
\midrule
All supplied records & 51 & 1.464--1.486--2.005 & 14, 26 & 8.04--8.87--21.43 & 0.16--0.19 \\
\bottomrule
\end{tabular}
\end{table}

\end{document}